\documentclass[journal=jacsat,manuscript=article]{achemso}

\usepackage[version=3]{mhchem} 
\usepackage{booktabs}

\usepackage{xcolor} 

\newcommand*\update[1]{\textcolor{black}{#1}}
\newcommand*\updateii[1]{\textcolor{black}{#1}}

\author{Samuel D. Tomlinson}
\affiliation{Department of Engineering, University of Cambridge, Cambridge, UK}
\alsoaffiliation
{Centre for Climate Repair, Cambridge, UK}
\alsoaffiliation
{School of Computing and Mathematical Sciences, University of Greenwich, London, UK}
\email{sdt50@cam.ac.uk}
\author{Aliki M. Tsopelakou}
\affiliation{Department of Engineering, University of Cambridge, Cambridge, UK}
\alsoaffiliation
{Centre for Climate Repair, Cambridge, UK}
\author{Tzia M. Onn}
\affiliation{Department of Engineering, University of Cambridge, Cambridge, UK}
\alsoaffiliation
{Centre for Climate Repair, Cambridge, UK}
\author{Steven R. H. Barrett}
\affiliation{Department of Engineering, University of Cambridge, Cambridge, UK}
\author{Adam M. Boies}
\affiliation{Department of Engineering, University of Cambridge, Cambridge, UK}
\alsoaffiliation
{Centre for Climate Repair, Cambridge, UK}
\altaffiliation
{Department of Mechanical Engineering, Stanford University, Stanford, USA}
\email{aboies@stanford.edu}
\author{Shaun D. Fitzgerald}
\affiliation{Department of Engineering, University of Cambridge, Cambridge, UK}
\alsoaffiliation
{Centre for Climate Repair, Cambridge, UK}
\email{sdf10@cam.ac.uk}

\title{Evaluating the scale of natural and mechanical airflows for surface-based atmospheric pollutant removal}

\abbreviations{IR,NMR,UV}

\begin{document}

\setlength{\baselineskip}{.8\baselineskip} 







\begin{abstract}
    \setlength{\baselineskip}{.8\baselineskip} 
    Removal strategies for atmospheric pollutants are increasingly being considered to mitigate global warming and improve public health. 
    However, the global potential of surface-based removal techniques (e.g. sorption, catalysis and filtration) has not yet been quantified based on the limits of pollutant transport.
    We evaluate the atmospheric pollutant transport to surfaces and assess the potential of surface-based removal technologies for deployment across the built environment, ventilation and convection systems, and the transportation fleet. 
    \updateii{Cities provide the highest transport-limited removal potential, with median annual atmospheric flow rates of 30 GtCO$_2$, 0.06 GtCH$_4$, 0.007 GtNO$_\text{x}$ and 0.0001 GtPM$_{2.5}$ to their total surface area. 
    Cities, solar farms and HVAC systems have flow rates large enough to potentially remove more than 1 GtCO$_2$/y (1 GtCO$_2$e/y for CH$_4$, 20-year GWP), if laboratory-scale removal efficiencies from the literature are applied to their total surface area, however, achieving this would require technological advances.} 
    \updateii{These values represent transport-limited upper bounds under current atmospheric concentrations; potential pollutant removal at these scales would alter atmospheric concentrations, thereby reducing the flow rate to surfaces.} 
    \update{Based on their transport-limited upper bounds, HVAC filters have the potential to achieve costs as low as \$600 per tCO$_2$ removed (\$2000 per tCO$_2$e) if CO$_2$-sorption (CH$_4$-catalyst) technologies are incorporated into their surfaces and performance is maintained through routine replacement, compared with \$3000 per tCO$_2$ (\$10000 per tCO$_2$e) for city surfaces, using literature values for these technologies’ material and application costs.} 
    \updateii{These estimates exclude 
    regeneration energy and electricity use associated with catalyst activation and the sorbent downstream chain, and therefore represent order-of-magnitude screening values.} 
    \update{These findings demonstrate that integrating surface-based pollutant removal technologies into infrastructure may offer a pathway to advance climate objectives, though further studies are needed to assess their feasibility in application, and application-implementation rates and cost. 
    }
\end{abstract}
\textbf{Synopsis:} \update{Surface-based removal technologies applied to existing infrastructure could enable atmospheric pollutant removal that supports climate goals. 
}

\vspace{-0.5cm}
\section{Introduction}

\update{Greenhouse gases (GHGs) such as carbon dioxide (CO$_2$), methane (CH$_4$) and nitrous oxide (N$_2$O) continue to rise at substantial rates, presenting sustainability and health challenges by accelerating climate change~\cite{kumar2021climate, jones2023national}.} 
In 2023, CO$_2$ levels surpassed 420 ppm, driven by fossil fuel use and deforestation, contributing to biodiversity loss and environmental degradation~\cite{prentice2001carbon, kumar2021climate}.
\update{Although CH$_4$ and N$_2$O are emitted in smaller quantities than CO$_2$, their 20-year global warming potentials (GWPs) are 84 and 273 times greater than CO$_2$, highlighting their disproportionate climate impacts despite lower atmospheric concentrations~\cite{kumar2021climate, jones2023national}.} 
\update{CO$_2$ is emitted from anthropogenic activities (e.g., fossil fuel combustion, land-use change) and natural processes (e.g., respiration, decomposition), and CH$_4$ and N$_2$O arise from a combination of anthropogenic sources, such as agriculture and fossil fuel processing, and natural microbial processes~\cite{balcombe2018methane, yusuf2012methane, forster2021earth, rowland2006stratospheric}. 
Many GHG sources are spatially distributed and occur at low concentrations, making them difficult to remove.} 
\update{In addition to GHGs, atmospheric pollutants such as particulate matter (PM), sulphur dioxide (SO$_2$), nitrogen oxides (NO$_\text{x}$) and ozone (O$_3$) contribute to a range of environmental and health risks~\cite{manisalidis2020environmental, world2021global}. 
PM, particularly particles smaller than 2.5 $\mu$m (PM$_{2.5}$), is a leading cause of respiratory and cardiovascular diseases, responsible for millions of premature deaths each year~\cite{sosa2017human, world2021global, polichetti2009effects}. 
PM is emitted from anthropogenic sources, such as fossil fuel combustion and processing, and natural sources, like wildfires. 
It can also form secondarily from precursor gases including SO$_2$ and NO$_\text{x}$, which have similar sources to primary PM and contribute to acid rain and ground-level O$_3$~\cite{cape2003ecological, amoatey2019emissions, boningari2016impact}. 
O$_3$, a secondary pollutant formed photochemically from NO$_\text{x}$ and volatile organic compounds (VOCs), exacerbates respiratory illness and contributes indirectly to radiative forcing, though its atmospheric lifetime is shorter than those of long-lived GHGs~\cite{zhang2019ozone}.} 

Given the diffuse nature of anthropogenic, biogenic and natural sources, coupled with the insufficient progress in emissions reductions, there is growing interest in atmospheric pollutant removal~\cite{prentice2001carbon, kumar2021climate, forster2021earth, rowland2006stratospheric, world2021global, polichetti2009effects, cape2003ecological, amoatey2019emissions, boningari2016impact}. 
\update{Atmospheric pollutant removal strategies can broadly be grouped into (i) surface-based approaches~\cite{sanz2016direct, luis2012recent, lunsford2000catalytic, guerrero2021methanotrophs, liu2020progress, parker1997electrostatic, zhao2022, Xu2009}, which capture or transform pollutants at solid interfaces, and (ii) airborne chemical methods~\cite{wuebbles2002atmospheric, randall2024cost}, such as radical generation or advanced photochemistry.} 
\update{While airborne methods show promise, this study focuses on surface-based removal methods such as sorption, catalysis and filtration, given that these approaches are more widely explored and developed.} 
Despite the interest in surface-based atmospheric pollutant removal~\cite{sanz2016direct, luis2012recent, lunsford2000catalytic, guerrero2021methanotrophs, liu2020progress, parker1997electrostatic, zhao2022, Xu2009}, the scale of potential removal remains unexplored. 
\update{This study quantifies the transport of CO$_2$, CH$_4$, NO$_\text{x}$ and PM$_{2.5}$ to natural and engineered surfaces to estimate upper-bound removal potentials, providing a comparative framework across environments rather than prescribing deployment pathways.} 

\updateii{Sorption technologies include liquid solvent-based systems (e.g., aqueous amines or hydroxide solutions) and solid sorbent-based systems (e.g., amine-functionalised materials), which capture CO$_2$ through chemical absorption or adsorption, as used, for example, in direct air capture (DAC)~\cite{sanz2016direct, custelcean2022direct}. 
For DAC, the removal efficiency is defined as the fraction of CO$_2$ removed from the airflow during passage through the capture device.
Reported values range from $10^{-3}$ to $10^{0}$ depending on the capture material, reactor configuration and operating conditions~\cite{soeherman2023overcoming}.}
However, this process requires substantial energy and involves high operational costs~\cite{mcqueen2020cost, mcqueen2021review}.
Membrane-based systems, which separate CO$_2$ through selectively permeable barriers, offer greater energy efficiency than DAC, but face challenges related to initial costs and scalability~\cite{luis2012recent}.
Hence, while both methods show promise, their widespread application has been limited.
In contrast, leveraging pre-existing natural and mechanical airflows, where the associated energy demands are already accounted for, may enable scalable and cost-effective surface-based methods for atmospheric pollutant removal.
\updateii{
Sorption-based systems require periodic regeneration and captured CO$_2$ requires concentration, compression, transport and sequestration. 
Each step imposes energy and cost constraints that determine whether surface-based approaches can achieve durable CO$_2$ removal. 
}

\updateii{Catalytic oxidation can convert CH$_4$ into CO$_2$, reducing radiative forcing rather than achieving net carbon removal, as CH$_4$ has a 20-year GWP approximately 84 times higher than CO$_2$~\cite{jones2023national}; however, performance optimisation across environments remains challenging~\cite{lunsford2000catalytic}.} 
Reported CH$_4$ removal efficiencies for thermal, photo and electrocatalysts vary widely, from $10^{-10}$ to $10^{0}$, reflecting differences across catalysts and experimental conditions~\cite{tsopelakou2024exploring, pennacchio2024physical}.
\updateii{Beyond catalyst materials and application, the feasibility of atmospheric pollutant removal technologies depends on fabrication, installation and energy inputs. 
For example, catalytic CH$_4$ oxidation may require thermal, photo or electrochemical activation and additional infrastructure to enable effective air--surface interaction, which can increase system-level costs and influence scalability across environments.}
Recent cost analyses suggest that using photocatalytic oxidation for atmospheric CH$_4$ removal is prohibitively expensive due to the high flow rates required for effective removal~\cite{randall2024cost, pennacchio2024physical, hickey2024economics}.
\updateii{Further analysis shows that targeting higher-concentration CH$_4$ sources can yield meaningful climate benefits at source concentrations above 10 ppm, with cost-effective operation using current technologies requiring concentrations above $10^{3}$ ppm~\cite{abernethy2023assessing}. 
Hydroxyl and chlorine radicals can oxidise CH$_4$ in the gas phase, and laboratory-scale photochemical systems have reported relatively low energy inputs for low-concentration pollutant removal~\cite{randall2024cost, krogsboll2025efficient}. 
}
Alternatively, atmospheric radicals have been suggested to break down CH$_4$ but at slow, condition-dependent rates~\cite{wuebbles2002atmospheric, randall2024cost} and methanotrophic bacteria can remove CH$_4$, albeit under controlled conditions~\cite{guerrero2021methanotrophs}.
This study examines catalytic oxidation as a strategy for atmospheric pollutant removal, exploring applications with varying surface fluxes and areas for scalable implementation. 

\update{High-efficiency particulate air (HEPA) filters are widely used for PM capture, though their performance depends on particle size distributions, pressure-drop constraints and replacement cycles~\cite{liu2020progress}; additional coatings could extend their removal potential to some gaseous pollutants. 
} 
Other methods for removing PM include electrostatic precipitators and adsorption materials such as activated carbon~\cite{parker1997electrostatic}.
These technologies offer removal efficiencies of $10^{-1}$ to $1$~\cite{schroth1996new}, but require periodic regeneration or replacement due to saturation~\cite{lowther2023factors}.
PM is also removed from the atmosphere through dry (e.g., gravitational settling, surface absorption) and wet deposition (e.g., scavenging by rain), with removal rates varying by region and environmental conditions~\cite{sippola2002particle, wu2018comparison, giardina2019atmospheric}.
While these mechanisms for PM removal are well-established, this study estimates how much PM$_{2.5}$ is removed from the atmosphere by filtration technologies and dry deposition, comparing these with the potential removal rates of other pollutants.
\update{Similar to CH$_4$, N$_2$O can be removed via catalytic reduction, which converts N$_2$O into nitrogen (N$_2$) and oxygen (O$_2$)~\cite{randall2024cost}.}
\updateii{Photochemical destruction approaches have also been investigated for N$_2$O and fluorinated anesthetic gases in controlled exhaust streams~\cite{rauchenwald2020new}.}
\update{Another surface-based removal process is biological denitrification, where bacteria reduce N$_2$O to N$_2$ under anaerobic conditions~\cite{tian2023global}. 
}
\update{SO$_2$, NO$_\text{x}$ and O$_3$ can be transformed through catalytic processes (e.g., selective catalytic reduction for NO$_\text{x}$, oxidation pathways for SO$_2$, photocatalysis for O$_3$), with variability in efficiency, by-products and technological maturity~\cite{zhao2022}.} 
Another removal method is scrubbing, where SO$_2$ and NO$_\text{x}$ are removed from gas streams by passing them through chemically reactive liquid solutions~\cite{Xu2009}.

In summary, although many surface-based pollutant removal methods exist (or are being proposed), they are often inefficient, energy-intensive and/or costly.
Furthermore, the literature lacks an assessment of the potential scale of reductions achievable with surface-based atmospheric pollutant removal, despite the need for scalable solutions.
\updateii{Surface-based pollutant removal in applications may operate in transport-, reaction- or regeneration-limited regimes~\cite{soeherman2023overcoming,tsopelakou2024exploring}. 
In transport-limited systems, removal is constrained by the atmospheric delivery of pollutants to the surface. 
However, removal can also be limited by reaction kinetics, catalyst activity at low concentrations, sorbent regeneration frequency, or application-specific parameters such as residence time, surface area and pressure drop. 
Here, using the transport flux, surface area and laboratory removal efficiency to estimate the potential removal provides an indicative estimate used to compare different applications, rather than evidence of transport-limited removal behaviour in specific applications.} 

In this study, we examine approaches that could utilise the substantial volumes of air transported in urban areas, heating, ventilation and air conditioning (HVAC) systems, and transportation networks.
Rather than focusing on the performance of specific surface-based technologies that remove specific pollutants, this study generally evaluates the scalability of atmospheric pollutant removal via surface-based interactions through transport limitations.
By applying representative laboratory-scale removal efficiencies from the literature, we estimate potential removal rates for sorption, catalysis and filtration across a range of environmental and flow conditions.
\updateii{Throughout this study, global-scale removal potentials are calculated assuming complete use of existing surfaces under current atmospheric concentrations representing transport-limited theoretical upper bounds rather than realistic deployment scenarios.} 
We compare the potential pollutant removal rates with intergovernmental targets and existing removal mechanisms, evaluating the scalability of these surface-based technologies.
\updateii{A screening analysis is performed to identify configurations where material and application costs alone may constrain large-scale deployment, recognising that full techno--economic feasibility depends on energy use and system integration.} 

\vspace{-0.5cm}
\section{Methods}

\update{The methods outlined in this section are used to calculate the pollutant flux or flow rate to the surface of different natural and mechanical environments. 
These fluxes or flow rates are independent of the removal technology and indicate the rate at which a given atmospheric pollutant could be delivered to the surface of a given environment (to potentially be removed). 
}
\updateii{In this work we distinguish between flow-over and flow-through transport regimes. 
For city surfaces, pollutant delivery is governed by boundary-layer transport, where flux to the surface depends on turbulent mixing and concentration gradients. 
In contrast, HVAC filtration operates as a flow-through configuration in which air is driven through a porous medium, and pollutant transport is governed by the velocity component normal to the medium and the concentration. 
External engineered systems such as aircraft represent a further regime, in which higher free-stream velocities generate thin turbulent boundary layers.} 
\updateii{
For gaseous species, transport to surfaces is governed primarily by advection and molecular or turbulent diffusion. 
For PM, transport may additionally involve gravitational settling, Brownian diffusion, inertial impaction and interception~\cite{hinds2022aerosol}. 
Pollutant-specific capture processes are incorporated separately where appropriate (e.g., through the diffusivity $D$ in~\eqref{eq:scaling_eq} and the HVAC filtration efficiency $E$ in~\eqref{eq:fibre_flow_rate}).} 

\vspace{-0.5cm}
\subsection{Scaling theory}

To characterise atmospheric pollutant transport dynamics across various systems, we define velocity ($u$, m/s), length ($l$, m) and surface area ($s$, m$^2$) scales specific to each flow application (e.g., cities, HVAC systems, transport). 
We also estimate the number of occurrences ($n$) for each flow type to ensure realistic scaling for global estimations. 
The limitations of this scaling approach and additional context are provided in Section S3. 
The thickness of the turbulent ($Re > 10^5$) boundary layer ($\delta$, m) is evaluated using $\delta = 0.4l/{Re}^{1/5}$, where the Reynolds number $Re = u l / \nu$~\cite{schlichting2016boundary} and $\nu$ is the kinematic viscosity (m$^2$/s).
\updateii{This empirical relation describes a turbulent boundary layer developing over a smooth flat plate under zero pressure gradient. 
In this work, it is used as an order-of-magnitude scaling for near-surface diffusive transport across environments, rather than as an exact representation of transport in urban flows, which are influenced by surface roughness and can be more appropriately described using parameters such as roughness length and displacement height~\cite{schlichting2016boundary}.} 

The normal pollutant flux $j_y$ (mol/(m$^2$s)) at the surface is determined by the diffusive flux, $-D c_y$, as the normal velocity at the surface is zero.
This normal pollutant flux depends on the diffusivity $D$ (m$^2$/s), concentration $c$ (mol/m$^3$) and the boundary layer thickness $\delta_c = \delta / {Sc}^{1/3}$ (m)~\cite{frank2017incropera}, where the Schmidt number ${Sc} = \nu / D$.
\updateii{Molecular diffusivities of gases in air are typically $O(10^{-5})$ m$^2$/s. 
In contrast, the Brownian diffusivity of particles decreases with particle diameter according to the Stokes--Einstein relation, with values around $O(10^{-8})$ m$^2$/s for 0.1~$\mu$m particles and $O(10^{-11})$ m$^2$/s for 1~$\mu$m particles~\cite{hinds2022aerosol}. 
Particle diffusivities are therefore several orders of magnitude smaller than gaseous diffusivities. 
}
An effective diffusivity $D_e$ (m$^2$/s) is included to account for enhanced turbulent mixing, which alters pollutant transport rates~\cite{pope2000}.
We will assume well-mixed concentrations and constant effective diffusivities throughout. 
Their spatio-temporal variations due to environmental factors and sources will be pursued in future work.

Using these scales, the streamwise flow rate of pollutants $Q$ (m$^3$/s) through the boundary layer of each application is given by
\begin{equation} \label{eq:scaling_Q}
    Q \approx n u \delta l.
\end{equation}
The normal flow rate of pollutants $q_y$ (mol/s) to the surface of each application is given by
\begin{equation} \label{eq:scaling_eq}
q_y \approx \frac{n s (D + D_e) c}{\delta / {Sc}^{1/3}}.
\end{equation}
\update{For some applications (e.g., cities, solar farms), multiple characteristic length scales are applied (e.g., buildings, solar panels), whereas for others (e.g., ducts, aeroplanes), only a single length scale is used. 
For applications with a single length scale, \eqref{eq:scaling_Q}--\eqref{eq:scaling_eq}  are evaluated using parameter distributions from Tables S1--S3. 
For multiple length scale applications, \eqref{eq:scaling_Q}--\eqref{eq:scaling_eq} are evaluated separately for each length scale and the results are combined. 
}

\vspace{-0.5cm}
\subsection{Empirical relationships} 

Empirical models can be used to link flow characteristics with the potential for pollutant removal~\cite{frank2017incropera}. 
These relationships have been applied in CH$_4$ and N$_2$O removal studies via photocatalysis~\cite{tsopelakou2024exploring, randall2024cost}.
These relationships are evaluated using $l$, $s$ and $u$, supplemented with measured values such as wall shear stresses ($\tau_w$ in kg/ms$^2$), mass transfer coefficients ($m_c$ in m/s) and heat transfer coefficients ($h_c$ in W/m$^2$K). 
A detailed discussion of these empirical relationships and their applications can be found in Section S3.
Analogies exist between key dimensionless numbers: the Nusselt number (${Nu} = h_c l / k$), the Sherwood number (${Sh} = m_c l / D$) and the drag coefficient ($C_d = \tau_w / (\rho u^2$)), where $k$ is the thermal conductivity (W/(mK)) and $\rho$ is the fluid density (kg/m$^3$)~\cite{frank2017incropera}.
These analogies allow for the formulation of the normal pollutant flow rate to the surface of each application as follows
\begin{equation} \label{eq:empirical}
q_y \approx 0.03 {Re}^{4/5} {Sc}^{1/3} n s D c/l.
\end{equation}
Following Tsopelakou \textit{et al.}~\cite{tsopelakou2024exploring}, we can use \eqref{eq:empirical} to estimate the streamwise flow rate through each application as
\begin{equation} \label{eq:aliki}
    Q \approx u A \approx l m_c P,  
\end{equation}
where $P$ is the wetted perimeter and $A$ is the cross-sectional area of the channel or boundary layer. 
\update{Eqs. \eqref{eq:empirical}--\eqref{eq:aliki}  are evaluated using the parameter distributions listed in Tabs. S1--S3. 
}

\vspace{-0.5cm}
\subsection{Energy, power and drag measurements} 

We now outline formulas to determine the flow rate based on energy, power and drag measurements. 
More details are given in Section S3.
For external flows over urban areas or solar farms, we establish a control volume with a length \( l \) and height \( \delta \).
In the case of uni-directional, steady-state flow over a surface with a heat flux \( \phi \) (W/m$^2$), energy conservation dictates that the energy entering the control volume must balance with the energy exiting it~\cite{frank2017incropera}.
This relationship can be expressed as
\begin{equation} \label{eq:nat}
    Q \approx \frac{n s \phi}{\rho c_p \Delta T }, \quad  \quad q_y \approx \frac{Q l (D+D_e) c}{\delta^2 u},
\end{equation}
where \( c_p \) (m$^2$/(s$^2$K)) is the specific heat capacity of air at constant pressure and \( \Delta T \) (K) is the temperature difference between the surface and the surrounding atmosphere.
The flow rates are related using the scaling theory discussed in Section S3, such that \( Q l (D+D_e) c / (\delta^2 u) \sim (u \delta) l (D+D_e) c / (\delta^2 u) \sim l (D+D_e) c / \delta \sim q_y \).

In the context of internal flow through HVAC systems, the total power consumed by the fans, denoted as \( P \) (kW), and the specific fan power (SFP) (W/(m$^3$/s)), can be utilised to evaluate the streamwise flow rate through the duct as follows
\begin{equation} \label{eq:int}
    Q \approx \frac{L P}{\text{SFP}},
\end{equation}
where \( L \) is the leakage factor~\cite{schild2009recommendations}.
For external airflows acting on transportation systems, the streamwise flow rate through the boundary layer can be estimated using the drag force acting on the aeroplane, train or automobile~\cite{white2003fluid}. 
This is expressed by the equation
\begin{equation} \label{eq:dragg}
    Q \approx \frac{2 n F_d}{\rho u C_d},
\end{equation}
where \( F_d \) (N) is the drag force.
\update{Eqs. \eqref{eq:nat}--\eqref{eq:dragg} are evaluated using the parameter distributions listed in Tabs. S1--S3. 
}

\vspace{-0.5cm}
\subsection{Industry standards}

For internal airflows through HVAC ducting, the streamwise flow rate can be evaluated using three estimates~\cite{ASHRAE2024}.
First, $Q$ can be expressed in terms of the cubic feet per minute per person (\( \text{CFM}_{\text{pp}} \), ft$^3$/min), given by
\begin{equation} \label{eq:cfm_pp}
    Q \approx L \text{CFM}_{\text{pp}} n_{\text{pp}},
\end{equation}
where \( n_{\text{pp}} \) (--) is the total number of occupants in the building. 
Second, $Q$ can be calculated based on the CFM per unit area (\( \text{CFM}_{\text{pm}^2} \), ft$^3$/min/m$^2$), described by the equation
\begin{equation}
    Q \approx L \text{CFM}_{\text{pm}^2} a,
\end{equation}
where \( a \) (m$^2$) is the total floor area of the building.
Finally, $Q$ can be assessed using the air changes per hour (ACH, s), given by
\begin{equation} \label{eq:ach}
    Q \approx L \text{ACH} v,
\end{equation}
where \( v \) (m$^3$) is the volume of air within the building.
\update{Eqs. \eqref{eq:cfm_pp}--\eqref{eq:ach} are evaluated using the parameter distributions listed in Tabs. S1--S3. 
}

\vspace{-0.5cm}
\subsection{Fully-developed profiles} 


For internal airflows within HVAC systems, we approximate solutions of the Navier-Stokes equations and boundary conditions~\cite{white2003fluid}.
The normal flow rate of pollutants to the channel walls is given by
\begin{equation} \label{eq:duct_law}
q_y \approx \frac{n s (D + D_e) c}{h},
\end{equation}
where $2h$ is the channel diameter.
For internal airflows through HVAC filters, we adjust the velocity from the duct flow (with velocity $u$, diameter $2h$ and cross-channel area $A$) to the pleat flow (with velocity $u_{\boldsymbol{n}}$, diameter $2h_p$, filter sheet area $A_p$, number $m$).
\update{Flow-through systems (e.g., HVAC filters) are treated as a distinct category of surface-based removal because air is forced through a reactive filter sheet, increasing pollutant--surface encounter frequency relative to diffusive exposure along surfaces. 
}
We apply the principle of mass conservation to match average velocities, leading to $A u \approx m A_p u_{\boldsymbol{n}}$.
The flow rate of pollutants to the fibre sheet is
\begin{equation} \label{eq:fibre_flow_rate}
    q_{\boldsymbol{n}} \approx n m A_p E u_{\boldsymbol{n}} c,
\end{equation}
where $E$ is the collection efficiency, related to the permeability and porosity of the filter medium~\cite{hinds2022aerosol}.
\updateii{Particle capture in HVAC filters is not governed by diffusion alone. 
Depending on particle size and flow conditions, dominant mechanisms include Brownian diffusion, interception and inertial impaction~\cite{hinds2022aerosol}. 
These mechanisms are incorporated into the collection efficiency term $E$ in~\eqref{eq:fibre_flow_rate}. 
For $\mu$m-scale particles typical of atmospheric PM$_{2.5}$, representing mass-based metrics, filtration is often inertia- or interception-dominated, such that the pollutant removal rate depends primarily on the flow-through velocity rather than molecular diffusivity. 
}
\update{Eqs. \eqref{eq:duct_law}--\eqref{eq:fibre_flow_rate} are evaluated using the parameter distributions listed in Tabs. S1--S3, and the resulting distribution is then combined with the other methods from this section. 
}

\vspace{-0.5cm}
\section{Box model}

\updateii{
To evaluate how surface-based pollutant removal could alter atmospheric concentrations, we consider a simple well-mixed atmospheric box model. %
We assume the atmospheric concentration $C = C(t)$ evolves according to
\begin{equation}
    \frac{dC}{dt} = \frac{E}{M_\mathrm{atm}} - (k_\mathrm{nat}+k_\mathrm{surf})C,
\end{equation}
where $E = E(t)$ is the emission rate, $M_\mathrm{atm}$ is the total atmospheric mass of the pollutant, $k_\mathrm{nat}$ is the first-order rate constant for natural removal processes and $k_\mathrm{surf}$ is the first-order rate constant for potential surface-based removal. %
Assuming constant emissions, the solution is given by
\begin{equation}
C=C_\mathrm{eq}+(C_0-C_\mathrm{eq})\exp(-(k_\mathrm{nat}+k_\mathrm{surf})t),
\end{equation}
where $C_\mathrm{eq}= E/(M_\mathrm{atm}(k_\mathrm{nat}+k_\mathrm{surf}))$ is the corresponding equilibrium concentration and $C(0)=C_0$. %
For CO$_2$, the present atmospheric burden is approximately 3200 GtCO$_2$, with anthropogenic emissions of approximately 40 GtCO$_2$/y~\cite{friedlingstein2022global}. %
The lifetime of atmospheric CO$_2$ is on the order of 50 to 100 years, giving a natural removal rate constant $k_\mathrm{nat}\approx 0.01$ to 0.02 $y^{-1}$. %
For CH$_4$, the present atmospheric burden is approximately 5.4 GtCH$_4$ and the atmospheric lifetime is approximately 9 to 12 years~\cite{saunois2019global}, giving a natural removal rate constant $k_\mathrm{nat}\approx 0.1$ $y^{-1}$. %
The potential surface-based removal rate constant is estimated as $k_\mathrm{surf} = R_\mathrm{surf}/M_\mathrm{atm}$, where $R_\mathrm{surf}$ is the potential removal rate for the pollutant removal technology. %
Tropospheric mixing times are approximately one year globally and weeks to months regionally, supporting the use of a well-mixed approximation as an order-of-magnitude analysis for CO$_2$ and CH$_4$; more detailed spatial predictions would require chemical transport modelling and are beyond the scope of this work. %
}

\vspace{-0.5cm}
\section{\update{Parameter estimation and uncertainty}}

\update{To account for variability in atmospheric pollutant transport, we estimated ranges for key parameters across natural, internal and external environments, including characteristic length, surface area, velocity, and total and produced numbers for each environment (Tab. S2--S4). 
Minimum, mean and maximum values were compiled from published literature and, where available, measurement datasets (e.g., wind speeds in cities~\cite{era5_single_levels}, Fig.~\ref{fig:0}E). 
For parameters with limited data, we adopted a representative value from the literature for the mean and estimated a plausible range, typically spanning approximately one order of magnitude, to reflect environmental variability and reporting uncertainty. 
Each parameter was represented as a triangular distribution (minimum–mean–maximum) to capture uncertainty, and this distribution was propagated through the atmospheric pollutant surface flux and flow rate equations \eqref{eq:scaling_Q}--\eqref{eq:fibre_flow_rate}. 
This generated a distribution of outcomes, which we illustrate in the Results using box-and-whisker plots, where the central line denotes the median, the box bounds represent the interquartile range, and the whiskers indicate the full range of simulated values. 
Mean values are overlaid for each method to aid comparison. 
}

\vspace{-0.5cm}
\section*{Results}

\begin{figure*}[t!]
\centering 
\includegraphics[width=\linewidth]{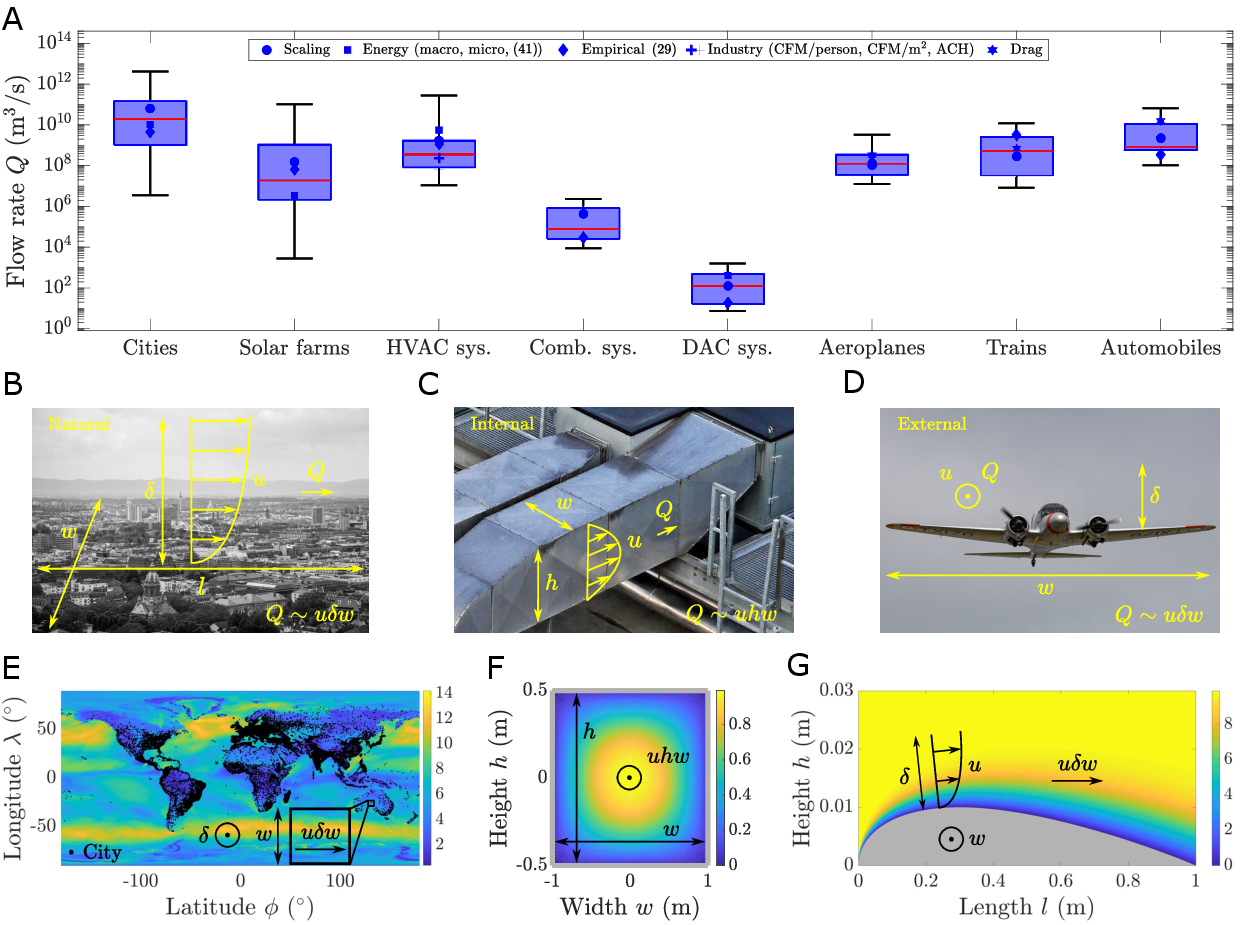}
\caption{
Flow rates through global environments.
(A) Streamwise flow rates ($Q$) for natural airflows (e.g., cities, solar farms), internal airflows (e.g., HVAC, combustion, DAC systems) and external airflows (e.g., aeroplanes, trains, automobiles), with symbols denoting the methods used to estimate flow rates.
\update{Box-and-whisker plots show the distribution of flow rates, where boxes represent the interquartile range (25th–75th percentiles), central lines denote medians, whiskers show the full range of values, and symbols indicate mean flow rates derived from different estimation methods.}
\update{Scaling \eqref{eq:scaling_Q}--\eqref{eq:scaling_eq}, energy \eqref{eq:nat}--\eqref{eq:int}, empirical \eqref{eq:empirical}--\eqref{eq:aliki}, industry \eqref{eq:cfm_pp}--\eqref{eq:ach} and drag \eqref{eq:dragg} approaches (using Tables S1--S3) are used to estimate $Q$. 
Schematics of the velocity field ($u$), lengthscales ($l$, $\delta$, $w$ and $h$) and flow rate for a (B) city, (C) HVAC system and (D) aeroplane.
\update{(E) Global wind speed distribution (m/s), with city data highlighted in black. 
(F) Streamwise velocity profile (m/s) through an HVAC system. 
(G) Boundary layer development (m/s) across an aerofoil surface. 
}
}
} 
\label{fig:0} 
\end{figure*}

\update{Natural (e.g., cities, solar farms), internal (e.g., HVAC, combustion, DAC systems) and external environments (e.g., aeroplanes, trains, automobiles) exhibit substantial airflow volumes through their boundary layers, with global medians of $2\times10^{10}$ m$^3$/s (range: $3\times10^{6}$ to $4\times10^{12}$ m$^3$/s) for natural, $4\times10^{8}$ m$^3$/s ($1\times10^{7}$ to $3\times10^{11}$ m$^3$/s) for internal and $8\times10^{8}$ m$^3$/s ($1\times10^{8}$ to $7\times10^{10}$ m$^3$/s) for external environments (Fig.~\ref{fig:0}A). 
Our analysis of variable parameters including length, surface area, velocity and number of environments, suggests that boundary-layer area and airflow velocity largely determine the flow rate through each environment: cities have the largest total surface area, whereas aircraft have higher velocities (Fig. 1B--G; Tabs. S2--S4). 
The parameter distributions define the range of flow-rate values and indicate the relative potential of these environments for practical deployment of removal technologies. 
}

\vspace{-0.25cm}
\subsection*{Surface fluxes of atmospheric pollutants} \label{sec:fluxes}

\subsubsection{\updateii{Natural flow over surfaces}}

\update{The following results describe the atmospheric pollutant fluxes to the surfaces of different environments; these values are determined solely by transport processes and are therefore also independent of the removal technology.}
Fig.~\ref{fig:3}A shows normalised atmospheric pollutant fluxes to the surfaces of natural environments (columns 1--2) in cities and solar farms. 
GHG fluxes exceed other pollutant fluxes due to higher diffusivities and atmospheric concentrations (Fig.~\ref{fig:3}C). 
\update{
Shorter development lengths reduce boundary layer thickness, increase turbulence intensity and enhance the transfer of atmospheric pollutants to surfaces~\cite{pope2000}. 
Cities exhibit average surface fluxes of $2\times10^{-8}$ mol/(m$^2$s) for CO$_2$ (range: $4\times10^{-10}$ to $1\times10^{-7}$ mol/(m$^2$s)), while solar farms exhibit average surface fluxes of $2\times10^{-7}$ mol/(m$^2$s) for CO$_2$ ($5\times10^{-9}$ to $1\times10^{-6}$ mol/(m$^2$s)) (Fig.~\ref{fig:3}C, Tab. S2). 
Although solar farms exhibit higher surface fluxes, the total surface area of cities will result in greater pollutant flow rates to their surfaces.} 
Using empirical formulae detailed in Methods, average atmospheric pollutant fluxes at the repeating unit scale (e.g., to the surface of individual buildings, $3\times10^{-2}$ m$^{-1}$) are larger than those at the combined unit scale (e.g., to the surface of entire cities, $7\times10^{-3}$ m$^{-1}$) due to steeper concentration gradients (Figs.~\ref{fig:3}A, D, E).
\update{Urban turbulence and building spacing will affect pollutant flow rates~\cite{britter2003flow}. 
In Methods, these effects are incorporated via turbulent diffusivity and boundary-layer formulations at city and building scales. 
}

\vspace{-0.25cm}
\subsubsection{\updateii{Internal flow over and flow through surfaces}}

Figs.~\ref{fig:3}A--B highlight the enhanced atmospheric pollutant transport to the surfaces of internal airflows (columns 3--5) in HVAC, combustion, DAC systems (Fig.~\ref{fig:3}A) and HVAC filters (Fig.~\ref{fig:3}B). 
Comparing HVAC, combustion and DAC systems, which show average surface fluxes of $7\times10^{-5}$, $2\times10^{-4}$ and $6\times10^{-5}$ mol/(m$^2$s) for CO$_2$ in Fig.~\ref{fig:3}C, demonstrates that higher airflow velocities enhance atmospheric pollutant transport.
The estimated value for DAC is comparable to $4\times10^{-5}$ mol/(m$^2$s) estimated in McQueen \textit{et al.}~\cite{mcqueen2021review}.
\updateii{Fluxes are compared on a per-unit-area basis (mol/(m$^{2}$s)) (Fig.~\ref{fig:3}), while total flow rates are calculated by multiplying these fluxes by the total surface area of each environment (Gt/y) (Fig.~\ref{fig:6}).} 
\update{Consistent with our definition of flow-over and flow-through systems (Methods), HVAC filters enhance encounter rates compared with passive duct walls.}
At the duct wall, the rate at which pollutants move perpendicular to the surface (normal flux) depends on diffusion and is proportional to $D c / h$ (e.g., an average of $7\times10^{-5}$ mol/(m$^2$s) for CO$_2$, Fig.~\ref{fig:3}C, F), where $D$ is the diffusivity of the pollutant, $c$ is its concentration and $h$ is the height of the channel.
In contrast, for HVAC filters, the pollutant flux through the porous fibre sheet is proportional to $u_{\boldsymbol{n}} c$ (e.g., an average of $3\times10^{-3}$ mol/(m$^2$s) for CO$_2$, Fig.~\ref{fig:3}C, G), where $u_{\boldsymbol{n}}$ is the airflow velocity through the fibre sheet.
Since $D$ and $c$ are typically small relative to $u_{\boldsymbol{n}}$ and $h$, $u_{\boldsymbol{n}}c$ is generally larger than $Dc/h$, highlighting the benefits of utilising advective over diffusive transport to improve atmospheric pollutant fluxes to surfaces.

\vspace{-0.25cm}
\subsubsection{\updateii{External flow over surfaces}}

\begin{figure*}[t!]
    \centering
    \includegraphics[width=\linewidth]{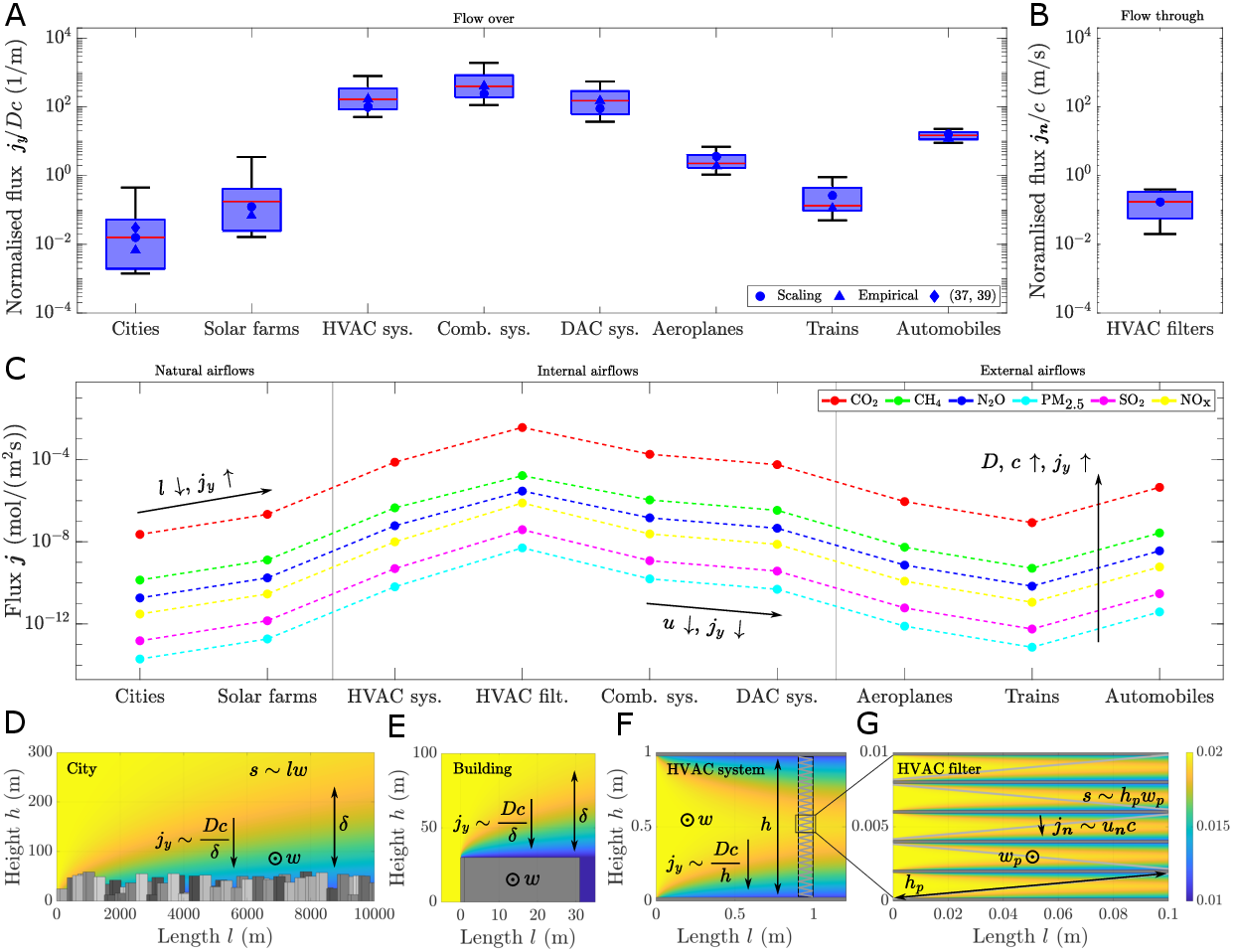}
    \caption{
    Atmospheric pollutant fluxes to the surfaces of global environments.
    (A) Atmospheric pollutant fluxes ($\boldsymbol{j}$), normalised by diffusivity ($D$) and concentration ($c$), to the surfaces of natural airflows (e.g., cities, solar farms), internal airflows (e.g., HVAC, combustion, DAC systems) and external airflows (e.g., aeroplanes, trains, automobiles).
    (B) Atmospheric pollutant fluxes, normalised by concentration, to the surfaces of HVAC filters.
    (C) Atmospheric pollutant fluxes to surfaces, averaged across the different methods, for CO$_2$, CH$_4$, N$_2$O, PM$_{2.5}$, SO$_2$ and NO$_\text{x}$.
    \update{Fluxes are estimated using scaling \eqref{eq:scaling_Q}--\eqref{eq:scaling_eq} and empirical \eqref{eq:empirical}--\eqref{eq:aliki} approaches, with parameter distributions from Tables S1--S3. 
    }
    Example concentration boundary layer, velocities ($u$ and $u_{\boldsymbol{n}}$), lengths ($l$, $\delta$, $w$, $h$, $w_p$, $h_p$) and normal pollutant flux to a (D) city surface (mol/m$^3$), (E) building surface (mol/m$^3$), (F) duct wall (mol/m$^3$) and (G) HVAC filter sheet (mol/m$^3$).
    }
    \label{fig:3}
\end{figure*}

Lastly, we evaluate the normalised atmospheric pollutant fluxes to the surfaces of external airflows over aeroplanes, trains and automobiles (columns 6--8, Fig.~\ref{fig:3}A).
Automobiles demonstrate higher pollutant fluxes to surfaces than trains and aeroplanes (e.g., an average of $4\times10^{-6}$, $8\times10^{-8}$, $9\times10^{-7}$ mol/(m$^2$s) for CO$_2$, Fig.~\ref{fig:3}C), due to smaller characteristic lengths. 
However, aeroplanes generate more turbulent transport due to their higher velocities, resulting in a pollutant flux that is approximately 23\% of that to automobile surfaces.

\vspace{-0.25cm}
\subsection*{Surface flow rates of atmospheric pollutants}

\subsubsection{Total surfaces}

\update{The atmospheric surface flow rates presented here are calculated from the fluxes and are independent of the removal method employed.
For comparison of total flow to surfaces across applications, we include both flow-over and flow-through categories, even though their underlying transport mechanisms differ. 
}
Fig.~\ref{fig:6} quantifies the flow rates of atmospheric CO$_2$, CH$_4$, PM$_{2.5}$ and NO$_\text{x}$ to all global surfaces.
These flow rates are evaluated by multiplying the fluxes from Fig.~\ref{fig:3} by the relevant surface area for each application ($s$ in Fig.~\ref{fig:3}D--G) and the number of each application.
\update{Natural and internal systems differ in the total pollutant flow to their surfaces, with cities channelling an average of 30 GtCO$_2$/y, three times higher than HVAC systems (10 GtCO$_2$/y) and nearly an order of magnitude above HVAC filters (4 GtCO$_2$/y) (Fig.~\ref{fig:6}A). 
While this indicates that city environments offer the largest surface area to reach climate-relevant theoretical removal bounds, realising this potential would require coating surfaces with removal technologies on a massive scale and maintaining their performance over decades. 
HVAC filters, although receiving lower pollutant flow rates to their surfaces, provide more controlled conditions that may simplify fitting, monitoring and maintenance; moreover, their regular replacement makes accessing the installed base more feasible, making them a potentially more promising near-term option. 
}
Conversely, external airflows and small-surface-area internal systems (columns 5--8) show lower atmospheric pollutant flow rates to their surface, e.g., an average of 0.1 GtCO$_2$/y for aeroplanes, 0.7 GtCO$_2$/y for trains, $7\times10^{-3}$ GtCO$_2$/y for combustion systems and $3\times10^{-7}$ GtCO$_2$/y for DAC systems, despite external airflows exhibiting higher atmospheric pollutant fluxes to their surfaces in Fig.~\ref{fig:3}A--C. 

\updateii{For context, the global ecosystem acts as a carbon sink of approximately 10 GtCO$_2$/y, primarily through photosynthesis and biomass accumulation, while upland soils remove roughly 0.03 GtCH$_4$/y through methanotrophic oxidation~\cite{friedlingstein2022global, dlugokencky2011global}. 
As shown in Fig.~\ref{fig:6}A, these ecosystem removal rates provide a benchmark for the transport-limited pollutant flow rates to surfaces estimated here. 
For reference, the global vegetation surface area is approximately $O(10^{14})$ m$^2$~\cite{fang2019overview}, two orders of magnitude larger than the median urban surface area considered here ($O(10^{12})$ m$^2$ estimated in Section S2). 
Therefore, although the median atmospheric delivery rates of CO$_2$ and CH$_4$ to urban surfaces slightly exceed these global ecosystem sink rates, the total realisable removal would likely be lower due to the smaller urban surface area and the dependence of removal on surface reactivity, residence time and energy constraints.} 

\vspace{-0.25cm}
\subsubsection{Produced surfaces}

\begin{figure*}[t!] 
\centering 
\includegraphics[width=.98\linewidth]{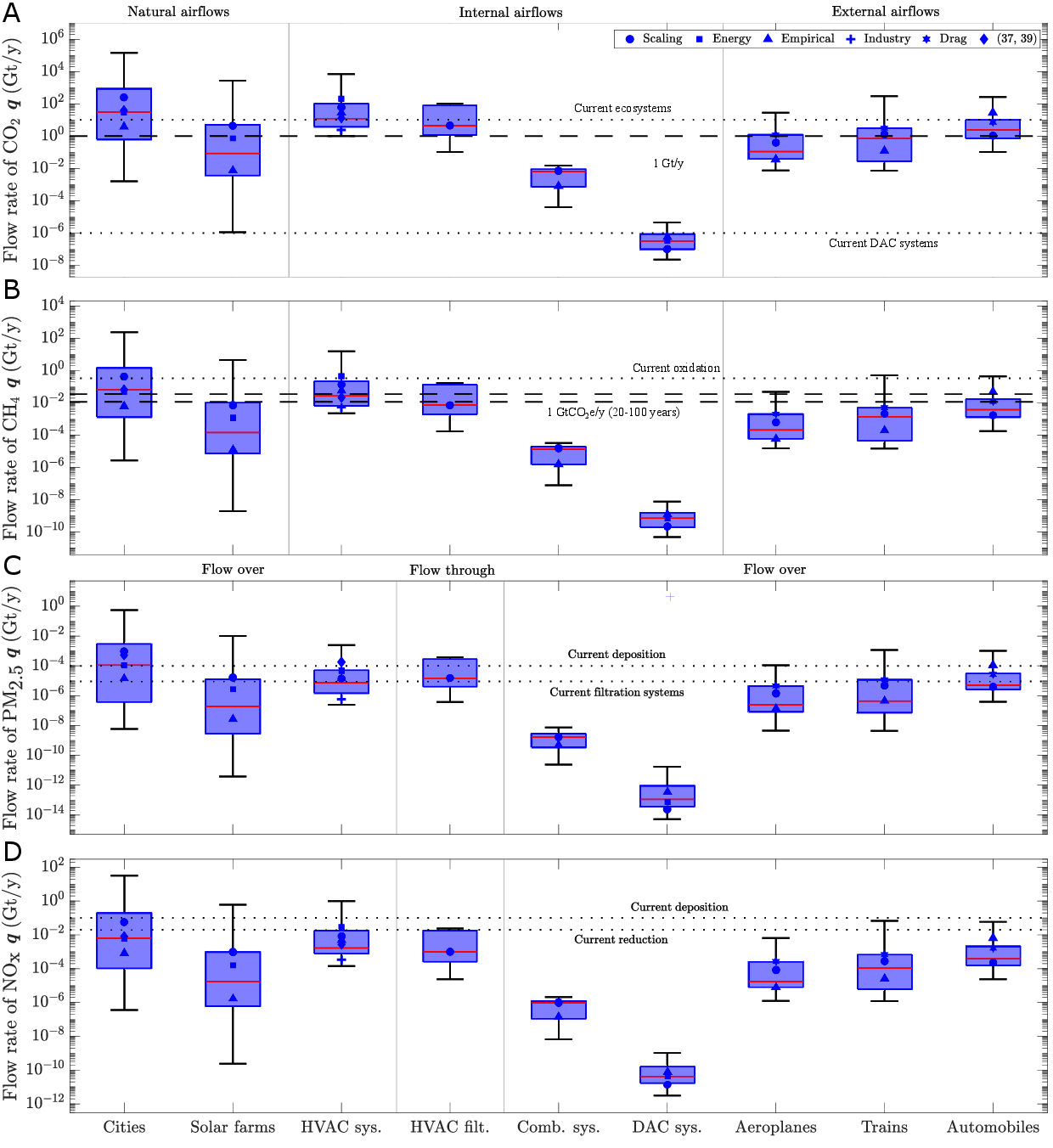} 
\caption{
Atmospheric pollutant flow rates to the surfaces of global environments. 
Flow rates ($\boldsymbol{q}$) of atmospheric (A) CO$_2$, (B) CH$_4$, (C) PM$_{2.5}$ and (D) NO$_\text{x}$ to the total existing surfaces in natural (e.g., cities, solar farms), internal (e.g., HVAC systems, HVAC filters, combustion systems, DAC systems) and external (e.g., aeroplanes, trains, automobiles) environments, with symbols representing the methods used to estimate flow rates.
Horizontal lines mark removal rates for atmospheric CO$_2$, CH$_4$, PM$_{2.5}$ and NO$_\text{x}$ due to ecosystems, oxidation, deposition and reduction~\cite{friedlingstein2022global, dlugokencky2011global, tian2023global, wu2018comparison, Dore2007, Xu2009, zhao2022}.
\update{Flow rates are derived using scaling \eqref{eq:scaling_Q}--\eqref{eq:scaling_eq}, energy \eqref{eq:nat}--\eqref{eq:int}, empirical \eqref{eq:empirical}--\eqref{eq:aliki}, industry \eqref{eq:cfm_pp}--\eqref{eq:ach} and drag \eqref{eq:dragg} approaches and Tables S1--S3. 
}
} 
\label{fig:6} 
\end{figure*}

\update{To approximate potential deployment rates, we scale atmospheric pollutant transport to yearly produced surfaces rather than total existing surfaces. 
Because infrastructure is replaced or expands gradually, flow rates to new surfaces decrease: CO$_2$ transport to urban surfaces drops from 30 Gt/y for existing infrastructure to 0.2 Gt/y for new surfaces, while solar farms decline from 0.08 Gt/y to 0.03 Gt/y (Fig.~\ref{fig:6}; Fig. S2). 
This contrast arises from differing growth rates: solar capacity is expanding at about 20\% per year, compared to approximately 4\% annual growth in urban area (Tab. S2). 
These results indicate that although cities contain the largest absolute removal capacity, solar farms accumulate removal potential more rapidly, underscoring that deployment timescales, not just total capacity, are critical for planning implementation. 
}

\vspace{-0.25cm}
\subsection*{Potential removal rates of atmospheric pollutants}

\subsubsection*{\update{CO$_2$ using sorption}}

\update{Having established the flow rate of CO$_2$ to the surfaces of various natural and mechanical environments, we assess the transport-limited upper bound to identify which environments could in principle remove atmospheric CO$_2$ from the atmosphere at rates exceeding the 1 GtCO$_2$/y benchmark~\cite{ipcc2023synthesis}.}  
\updateii{The feasibility of applying surface-based technologies in these environments, including potential impacts on their primary function, is discussed further in the Section S5.}
\update{The atmospheric CO$_2$ flow rate exceeds 1 GtCO$_2$/y for cities (based on the IQR: $6\times10^{-1}$ to $9\times10^{2}$ GtCO$_2$/y), solar farms ($4\times10^{-3}$ to $5\times10^{0}$ GtCO$_2$/y), HVAC systems ($4\times10^{0}$ to $1\times10^{2}$ GtCO$_2$/y), filters ($1\times10^{0}$ to $8\times10^{1}$ GtCO$_2$/y), aeroplanes ($4\times10^{-2}$ to $1\times10^{0}$ GtCO$_2$/y), trains ($3\times10^{-2}$ to $3\times10^{0}$ GtCO$_2$/y) and automobiles ($7\times10^{-1}$ to $1\times10^{1}$ GtCO$_2$/y), confirming that for these environments atmospheric transport is not the binding constraint relative to the 1 GtCO$_2$/y benchmark (Fig.~\ref{fig:6}A).}  
\updateii{
However, achieving this theoretical upper bound would require coating surface areas on the order of $10^{12}$ m$^2$, implying a very large global-scale material production and long-term maintenance effort.}
\update{If one only considers flow rates to annually produced surfaces (Fig. S2A), which can provide an estimate for potential deployment rate or market penetration, feasible applications narrow to cities, solar farms, HVAC systems and filters. 
For example, if sorbents were applied only to new building materials, the transition from annually produced surfaces to total existing surfaces would occur gradually over decades (Tab. S2). 
}

The realization of the above transport-limited potentials is constrained by the fact that removal efficiencies are often suboptimal, and the deployment of sorbents across large infrastructure surface areas is limited by the necessity of a complete removal chain.
\update{For example, DAC systems using sorbents exhibit removal efficiencies of around \(5 \times 10^{-2} \) (based on Soeherman et al.~\cite{soeherman2023overcoming}, which reviews various CO$_2$-removal technologies), giving a median removal potential of $2\times10^{0}$ GtCO$_2$/y for cities, $6\times10^{-1}$ GtCO$_2$/y for HVAC systems and $1\times10^{-1}$ GtCO$_2$/y for automobiles with sorbent surfaces.
} 
\updateii{If applications are reaction- or regeneration-limited, removal efficiencies may be lower. 
For example, if removal efficiency decreases by two orders of magnitude to $5\times10^{-4}$ in natural and external airflows, the corresponding removal potentials decrease to approximately $2\times10^{-2}$ GtCO$_2$/y for cities and $1\times10^{-3}$ GtCO$_2$/y for automobiles.} 
\update{We can compare these potential removal rates with current DAC systems which are removing around $10^3$ tCO$_2$/y (based on McQueen et al.~\cite{mcqueen2021review}, which reviews DAC approaches in the US) or $10^{-6}$ GtCO$_2$/y (Fig.~\ref{fig:6}A).} 

\updateii{For amine-functionalised solid sorbents used in DAC systems, typical working capacities are 1 to 3 mmol/g with regeneration cycles of approximately 15 to 85 min~\cite{mcqueen2021review}.
Total system energy requirements, including sorbent regeneration and air handling, are typically 6 to 10 GJ per tCO$_2$ removed. %
Scaling these operational requirements to the median sorbent removal potential estimated here (2 GtCO$_2$/y for city surfaces) implies energy inputs on the order of 12 to 20 EJ/y, corresponding to several percent of current global energy consumption. %
Given these constraints, realised net removal would be substantially lower than the theoretical flow rate estimates, depending on sorbent cycling efficiency, energy availability and proximity to sequestration infrastructure.}
\updateii{A potential removal rate of 2 GtCO$_2$/y, proportional to the predicted median sorbent removal potential for city surfaces, corresponds to a potential surface removal rate constant $k_\mathrm{surf}\approx 6\times10^{-4}$ $y^{-1}$ relative to the atmospheric burden of approximately 3200 GtCO$_2$~\cite{friedlingstein2022global}. %
This is small relative to the natural removal rate constant (0.01 to 0.02 $y^{-1}$), indicating that removal at this scale would have a limited effect on atmospheric CO$_2$ concentrations at the global scale. %
In contrast, the transport-limited upper bound removal potential of approximately 30 GtCO$_2$/y, corresponds to $k_\mathrm{surf}\approx 9\times10^{-3}$ $y^{-1}$, approaching the lower bound of the natural removal rate constant.  %
At this scale, potential removal could perturb atmospheric concentrations, reducing concentration gradients driving surface flux and changing the transport-limited estimates over time. %
Achieving removal at this scale would require the entire urban exterior surface area to be coated with sorbent materials operating at sustained  laboratory-scale efficiency, with each surface periodically regenerated and the captured CO$_2$ concentrated, compressed and sequestered --- a chain of requirements that currently lies well beyond demonstrated technological and economic feasibility.
}

\vspace{-0.25cm}
\subsubsection*{\update{CH$_4$ using catalysis}}

\update{Methane removal is more challenging due to its lower atmospheric concentration, but catalytic approaches can reach removal efficiencies of around \(4 \times10^{-2}\) (based on Tsopelakou et al.~\cite{tsopelakou2024exploring}, which reviews CH$_4$ photocatalyst data). 
When this laboratory-scale removal efficiency is applied to the upper range of modelled CH$_4$ flow rates (upper 5\% of estimated values), the resulting transport-limited upper-bound removal potentials reach 10 and 1 GtCH$_4$/y for cities and HVAC systems respectively (Fig.~\ref{fig:6}B), exceeding the estimated atmospheric oxidation rate of approximately
0.5 GtCH$_4$~\cite{dlugokencky2011global}, which is comparable to global CH$_4$ emission rates of around 0.55 GtCH$_4$/y~\cite{friedlingstein2022global}. 
\updateii{At ambient temperatures and concentrations, CH$_4$ oxidation can be kinetics-limited and removal efficiencies may be lower than laboratory values in some applications. 
If effective removal efficiencies in cities decrease to $4\times10^{-4}$, for example, the removal potentials would decrease to approximately $1\times10^{-1}$ GtCH$_4$/y, again illustrating the dependence of removal on application-specific conditions. 
}
Real-world constraints have varied pollutant reductions in field demonstrations of photocatalytic surfaces~\cite{Russell2021Review, ma2025}, which operate at much smaller scales than the global surfaces considered here.} 

\updateii{For CH$_4$, the transport-limited upper-bound potential removal rate for city surfaces of approximately 0.1 GtCH$_4$/y corresponds to $k_\mathrm{surf}\approx 0.02$ $y^{-1}$ relative to the atmospheric burden of approximately 5.4 GtCH$_4$~\cite{saunois2019global}. %
This is smaller than the natural atmospheric oxidation rate constant of approximately 0.1 $y^{-1}$, indicating that surface-based removal would have a limited effect on atmospheric concentrations under even these optimistic assumptions. %
Furthermore, this upper bound assumes sustained catalytic activity across all city surfaces at laboratory-scale catalytic efficiency.
In practice, CH$_4$ oxidation can be kinetics-limited at ambient concentrations~\cite{pennacchio2024physical}, and catalyst stability over deployment timescales under ambient conditions in applications if far from demonstrated. 
}

\updateii{Reported reaction probabilities for CH$_4$ oxidation on catalytic surfaces are typically $\gamma = O(10^{-9})$ per molecule--surface collision under ambient conditions~\cite{pennacchio2024physical}, such that removal efficiency scales as $\gamma N_\text{coll}$, where $N_\text{coll}$ is the number of molecule--surface collisions per transit. 
A laboratory removal efficiency of $4\times10^{-2}$ is therefore consistent with $N_\text{coll} = O(10^{7})$ collisions in reactor geometries~\cite{tsopelakou2024exploring}.  
In applications, $N_\text{coll}$ is expected to be several orders of magnitude lower due to shorter residence times, although flow-through configurations (e.g., HVAC filters) could increase molecule--surface interactions relative to flow-over surfaces. 
Photocatalytic CH$_4$ oxidation is further limited by low quantum efficiency, implying high photon flux requirements and increased energy demand~\cite{abernethy2023assessing}. 
In HVAC systems, UV illumination would generally be required for activation, introducing an energy cost, although UV irradiation is already used in some systems for microbial control~\cite{thornton2022impact}; outdoor systems may instead utilise sunlight, but remain constrained by low quantum efficiency and ambient CH$_4$ concentrations~\cite{randall2024cost}.
In the contexts considered here, removal would leverage existing airflow and energy sources (e.g., sunlight outdoors or UV in HVAC systems), so additional energy inputs are minimal and climate break-even is achievable.
If artificial airflow or irradiation were required, net climate benefit could be negative.
Moreover, all applications exceed \$100 per tCO$_2$e (see Tab.~\ref{tab:cost}), so none achieve cost break-even at current carbon prices.
Consistent with these constraints, prior studies report that photocatalytic CH$_4$ removal is generally cost-ineffective and may be climate-detrimental at ambient concentrations~\cite{abernethy2023assessing}. 
Removal efficiency, energy balance and cost per tonne are therefore strongly application-specific; the present analysis instead provides an upper bound on atmospheric transport to surfaces and does not imply technical, cost or ultimate viability.
} 

\vspace{-0.25cm}
\subsubsection*{\update{PM$_{2.5}$ using filtration and oxidised forms of NO$_\text{x}$ via deposition}}

\update{PM$_{2.5}$ is captured by HVAC filters which achieve removal efficiencies of around \(9.9\times10^{-1}\) (based on Schroth~\cite{schroth1996new}, which compares various filter classes), such that the average removal potential is $10^{-5}$ GtPM$_{2.5}$/y (Fig.~\ref{fig:6}C). 
Atmospheric removal rates approach $9\times10^{-6}$ GtPM$_{2.5}$/y (Fig.~\ref{fig:6}C), with estimates of 100 million filters processing 100 m$^3$/h at 99\% removal efficiency (based on Lowther et al.~\cite{lowther2023factors}, which compares PM removal across rooms in China). 
These removal efficiencies represent realistic operational averages that depend on filter replacement cycles and are higher than those of sorption or catalyst technologies. 
By contrast, PM$_{2.5}$, as well as the oxidised forms of NO$_\text{x}$, are captured when they interact with surfaces through deposition. 
As shown in Fig.~\ref{fig:6}C--D, dry deposition rates of $1\times10^{-4}$ GtPM$_{2.5}$/y and $1\times10^{-1}$ GtNO$_\text{x}$/y align with atmospheric pollutant flow rate estimates to urban surfaces (based on Wu et al.\cite{wu2018comparison} which measured dry and wet deposition of particulate matter during summer in Beijing and Dore et al.\cite{Dore2007} which modelled the atmospheric transport and deposition of sulphur and nitrogen across the UK), contributing around 1--10\% of global deposition.
}

\updateii{Air pollution is one of the leading environmental risk factors for premature mortality, with ambient air pollution contributing to approximately 4.2 million premature deaths annually and household air pollution contributing approximately 2.9 million, combining to approximately 7 million deaths per year worldwide~\cite{world2021global}. 
Global anthropogenic primary carbonaceous aerosol emissions in 2017 were approximately 22 Tg/y, representing a lower bound on total PM$_{2.5}$ emissions, and NO$_\text{x}$ emissions were approximately 122 Tg/y~\cite{mcduffie2020global}. 
The estimated median potential removal rates of $1\times10^{-4}$ GtPM$_{2.5}$/y and $2\times10^{-4}$ GtNO$_\text{x}$/y correspond to approximately 0.45\% and 0.16\% of these global anthropogenic emission totals, respectively. 
Assuming steady-state conditions and, as a first-order approximation, that concentration scales proportionally with emissions, these correspond to concentration reductions of approximately 0.11 $\mu$g/m$^{3}$ for PM$_{2.5}$ and 0.04 $\mu$g/m$^{3}$ for NO$_2$ at a typical urban mean concentration of 25 $\mu$g/m$^{3}$ for each pollutant.
Using exposure--mortality relationships of 2.8\% per 10 $\mu$g/m$^{3}$ for PM$_{2.5}$~\cite{kloog2013long} and 4\% per 10 $\mu$g/m$^{3}$ for NO$_2$~\cite{hamra2015lung}, these concentration reductions correspond to changes in pollutant-associated mortality of approximately 0.03\% and 0.02\%, respectively. 
These estimates are simplified and do not account for atmospheric chemistry, spatial heterogeneity or secondary aerosol formation; nonetheless, they imply that potential removal at the scales estimated here would correspond to generally small changes in pollutant-associated mortality at the global scale.} 

\vspace{-0.25cm}
\subsection*{Material and application costs}

\subsubsection*{\update{CO$_2$ using sorption and CH$_4$ using catalysis via literature estimates}}

\update{Finally, we estimate the potential material and application cost of scaling CO$_2$-sorption and CH$_4$-catalyst technologies, highlighting their capacity for atmospheric removal if deployed in different environments. 
These technologies are selected as promising approaches for GHG removal, with reported material and application costs ranging from \$16 to \$50 per m$^2$ for CO$_2$ sorbents (based on values given in Fig. 28 of Sanz et al.~\cite{sanz2016direct} and Fig. 7 of Kulkarni et al.~\cite{kulkarni2012analysis}) and \$15 to \$43 per m$^2$ for CH$_4$ catalysts (based on values given in the SI of Randall et al.~\cite{randall2024cost} and Hickey et al.~\cite{hickey2024economics}).} 
\updateii{These values exclude maintenance, regeneration energy, replacement frequency, performance degradation and operational energy inputs (e.g., UV illumination for photocatalysts, which may dominate system-level costs in internal environments, whereas increased fan power due to pressure-drop penalties is expected to be small), all of which could increase system-level costs. 
Total costs will vary with the application, and a techno--economic analysis would be required for each deployment scenario. 
Accordingly, the values reported here should be interpreted as order-of-magnitude screening estimates rather than total system costs.
}
\updateii{The material and application costs of \$16 to \$50 per m$^{2}$ for CO$_2$ sorbents exclude the infrastructure required for the capture-to-storage chain. %
Capture and regeneration costs reported for DAC systems typically range from \$88 to \$600 per tCO$_2$ removed, which is less than our lower bound estimate in Tab.~\ref{tab:cost}, while compression and sequestration can add an additional \$15 to \$21 per tCO$_2$, excluding transport costs which are site-specific~\cite{NASEM2019}. %
Applying these cost ranges to the capture potentials estimated here implies additional system costs of around \$103 to \$621 per tCO$_2$ potentially removed, or annual expenditures on the order of \$2$\times$10$^{11}$ to \$1.2$\times$10$^{12}$ per year if potential removal were sustained at 2~GtCO$_2$/y, which is comparable to around 1\% of current global GDP.%
}

\updateii{The costs of \$15 to \$43 per m$^{2}$ reported here for CH$_4$ oxidation catalysts include materials and application, with materials representing a smaller fraction of the total~\cite{randall2024cost}. %
Periodic maintenance adds approximately 11\% to the cost annually. 
In contrast, operational energy inputs can dominate system costs depending on the activation mechanism. 
For outdoor environments such as cities or transportation infrastructure, photocatalytic activation may be driven by natural radiation and therefore require little additional energy input. %
For enclosed environments such as HVAC systems, UV illumination may be required at irradiance levels of approximately 10 to 100 W/m$^{2}$~\cite{tsopelakou2024exploring}.
If applied across the global HVAC ventilation median surface area (approximately $5\times10^{11}$ m$^{2}$ estimated in Section S2), the required electrical power would be approximately 5 to 50 TW, corresponding to an annual electricity demand of approximately $4\times10^{4}$ to $4\times10^{5}$ TWh/y, which exceeds current global electricity production. %
At an electricity cost of \$0.08 per kWh, this corresponds to an annual electricity expenditure of \$3.5 to \$35 trillion, or approximately \$2100 to \$21000 per tCO$_2$e potentially removed (20-year GWP) at the median potential removal rate of $2\times10^{-2}$ GtCH$_4$/y, greater than the lower bound system cost of approximately \$2000 per tCO$_2$e estimated from materials and application alone in Tab.~\ref{tab:cost} 
However, in some indoor environments, UV illumination is already deployed for air treatment (e.g., microbial control), and photocatalytic systems could in principle leverage this existing infrastructure depending on wavelength compatibility and required irradiance, reducing installation and energy costs further.
This highlights that energy inputs can potentially become a dominant cost component for indoor configurations, and that the feasibility of catalytic CH$_4$ removal is sensitive to the application and whether artificial UV illumination is required.%
}

\renewcommand{\arraystretch}{1.4}
\begin{table*}[t!]
    \centering
    \caption{\update{Material and application cost of surface-based atmospheric pollutant removal in global environments, where values in brackets represent minimum--maximum ranges.}
    The upper-bound potential removal rate of atmospheric CO$_2$ and CH$_4$ per year using representative laboratory-scale removal efficiencies of $5\times{10}^{-2}$ for CO$_2$ sorbents and $4\times{10}^{-2}$ for CH$_4$ catalysis, surface area produced per year, technology material and application cost per year based on material and application cost per square meter~\cite{sanz2016direct, mcqueen2020cost, randall2024cost, hickey2024economics}, material and application cost per tonne of atmospheric CO$_2$ and CH$_4$ removed and material and application cost per tonne of CO$_2$e removed for different natural and internal flows. %
    Data for these natural and internal flows are divided into cities, solar farms, HVAC systems and filters that have potential removal rates of atmospheric CO$_2$ and CH$_4$ that exceed the 1 GtCO$_2$e/y target (20-year GWP) in Fig.~\ref{fig:6}.}
    \vspace{0.25cm}
    \resizebox{\textwidth}{!}{
    \begin{tabular}{@{}lcccccc@{}}
    \toprule
    \multicolumn{6}{c}{Carbon dioxide (CO$_2$)} \\ 
    \cmidrule(lr){2-6}
    Environment & Rem. rate (t/y) & Surf. area (m$^2$/y) & Tech. cost (\$/y) & Rem. cost (\$/t) & Rem. cost (\$/tCO$_2$e) \\ 
    \midrule
    Cities        & $[5\times10^2,\,2\times10^{10}]$ & $[5\times10^{9},\,1\times10^{12}]$ & $[8\times10^{10},\,5\times10^{13}]$ & $[3\times10^3,\,2\times10^8]$ & $[3\times10^3,\,2\times10^8]$ \\ 
    Solar farms   & $[2\times10^2,\,1\times10^{10}]$ & $[5\times10^7,\,3\times10^{11}]$   & $[8\times10^{8},\,1\times10^{13}]$  & $[1\times10^3,\,4\times10^6]$ & $[1\times10^3,\,4\times10^6]$ \\
    HVAC systems  & $[5\times10^5,\,5\times10^{9}]$ & $[3\times10^{9},\,6\times10^{10}]$ & $[5\times10^{10},\,3\times10^{12}]$ & $[6\times10^2,\,1\times10^5]$ & $[6\times10^2,\,1\times10^5]$ \\
    HVAC filters  & $[5\times10^5,\,1\times10^{9}]$ & $[2\times10^7,\,8\times10^{10}]$   & $[3\times10^{8},\,4\times10^{12}]$  & $[6\times10^2,\,4\times10^3]$ & $[6\times10^2,\,4\times10^3]$ \\
    \bottomrule
    \toprule
    \multicolumn{6}{c}{Methane (CH$_4$)} \\ 
    \cmidrule(lr){2-6}
    Environment & Rem. rate (t/y) & Surf. area (m$^2$/y) & Tech. cost (\$/y) & Rem. cost (\$/t) & Rem. cost (\$/tCO$_2$e) \\ 
    \midrule
    Cities        & $[2\times10^0,\,4\times10^7]$ & $[5\times10^{9},\,1\times10^{12}]$ & $[8\times10^{10},\,4\times10^{13}]$ & $[1\times10^6,\,4\times10^{10}]$ & $[1\times10^4,\,5\times10^8]$ \\
    Solar farms   & $[8\times10^{-1},\,4\times10^7]$ & $[5\times10^7,\,3\times10^{11}]$   & $[8\times10^{8},\,1\times10^{13}]$  & $[3\times10^5,\,1\times10^9]$ & $[4\times10^3,\,1\times10^7]$ \\
    HVAC systems  & $[4\times10^3,\,2\times10^7]$ & $[3\times10^{9},\,6\times10^{10}]$ & $[4\times10^{10},\,3\times10^{12}]$ & $[2\times10^5,\,1\times10^7]$ & $[2\times10^3,\,1\times10^5]$ \\
    HVAC filters  & $[2\times10^3,\,2\times10^6]$ & $[2\times10^7,\,8\times10^{10}]$   & $[3\times10^{8},\,4\times10^{12}]$  & $[2\times10^5,\,2\times10^6]$ & $[2\times10^3,\,2\times10^4]$ \\
    \bottomrule
    \end{tabular}}
    \label{tab:cost}
\end{table*}


\update{To evaluate cost-efficiency, we compare these material and application cost estimates with a benchmark of \$100 per tCO$_2$e potentially removed (20-year GWP), in line with recommended targets for GHG-removal technologies~\cite{ipcc2023synthesis}.} 
\updateii{We adopt the 20-year GWP to provide an upper-bound estimate of CO$_2$-equivalent benefit per tonne CH$_4$ potentially removed, consistent with the upper-bound transport-limited framework of this study, noting that CH$_4$ has an atmospheric lifetime of approximately 12 years and its warming impact is concentrated in the first few decades after emission. 
Sensitivity to alternative climate metrics is discussed in Section S8.
}
\update{The rationale for this is that if the material and application cost alone approaches or exceeds the targeted installed cost, this should be flagged, as it may indicate that further investigation has limited value.
We assume a removal efficiency of \(5\times10^{-2}\) for CO$_2$ sorbents~\cite{soeherman2023overcoming} and \(4\times10^{-2}\) for CH$_4$ catalysts~\cite{tsopelakou2024exploring}. 
For each application in Tab.~\ref{tab:cost}, we calculate the material and application cost per tonne of CO$_2$e removed by dividing the estimated material cost of sorbent or catalyst technologies by the corresponding CO$_2$ or CH$_4$ removal rate.} 
\updateii{These estimates assume a single installation cycle and exclude costs associated with replacement frequency (discussed further in Section S5), regeneration energy and energy penalties. 
The reported values should therefore be interpreted as order-of-magnitude materials and application screening estimates rather than full lifecycle costs. 
}
\update{All estimates across natural and internal environments exceed the \$100 per tCO$_2$e removed target. 
However, HVAC filters achieve potential material and application costs as low as \$600 per tCO$_2$ removed (range: \$$6\times10^{2}$ to $4\times10^{3}$ per tCO$_2$ removed) for sorption, and as low as \$2000 per tCO$_2$e removed (\$$2\times10^{3}$ to $2\times10^{4}$ per tCO$_2$e removed) for CH$_4$-catalyst technologies (Tab.~\ref{tab:cost}).} 
\update{This relatively lower material and application cost arises from the high pollutant flux to HVAC filter surfaces (e.g., an average of $0.003$ mol/(m$^2$s) for CO$_2$, Fig.~\ref{fig:3}C), which reduces the surface area requirement per unit of pollutant removal.}
\updateii{The feasibility of integrating removal technologies with periodic replacement and its effect on material and application cost is discussed further in Section S5.}

\vspace{-0.5cm}
\section*{Discussion} \label{sec:discussion}

\update{The potential for removing low concentrations of atmospheric pollutants, as a mitigation strategy for climate change and public-health crises, has yet to be fully explored. 
Surface-based removal technologies have been proposed as a potentially scalable approach, with examples including sorbents for CO$_2$ capture~\cite{mcqueen2021review}, catalysts to reduce VOCs and NO$_\text{x}$~\cite{shah2019review}, and HEPA filters for PM removal~\cite{liu2020progress}.} 
\updateii{
Viable CO$_2$ sorbent materials have been demonstrated at laboratory or pilot scale, for example in DAC~\cite{mcqueen2021review}, sorbent filters within ventilation systems~\cite{wu2025distributed} or as carbonation coatings applied to building surfaces~\cite{peng2023development}, and could in principle be integrated into wider infrastructure. 
Unlike carbonation coatings, in which captured CO$_2$ is mineralised at the surface for a single application, sorbent filters must undergo periodic regeneration, and the captured CO$_2$ must then be collected, compressed, transported and sequestered to achieve climate benefits.} 
\updateii{
Thus, while our analysis identifies large theoretical capture potentials based on atmospheric transport to surfaces, realising sustained removal at scale requires integration of regeneration, energy, transport and storage infrastructure, and is therefore constrained by system-level energy and economic considerations. 
The flow rate estimates reported here should accordingly be understood as upper bounds on what atmospheric transport and surface chemistry may permit, not as projections of achievable removal.
}

\updateii{Recent field deployments show photocatalytic systems achieving NO$_\text{x}$ removal of up to 66\% at street level and 82\% on wall surfaces under ambient conditions. In tunnel environments, reductions of up to 23\% have been reported under a light intensity of approximately 20 W/m$^{2}$, air velocities of approximately 0.4 m/s and relative humidity of 40 to 70\%~\cite{Russell2021Review}.} 
\updateii{These values represent near-surface removal efficiencies and are used here to define an upper-limit removal potential within a transport-based framework. 
However, Russell et al.~\cite{Russell2021Review}, conclude that outdoor deployments of TiO$_2$ coatings often yield ambient concentration reductions of around 2\% in the intermediate vicinity of treated surfaces. 
In addition, formation of secondary species has been observed in some studies, such that air-quality impacts depend on surface reactivity and atmospheric chemistry.} 
\update{Ma et al.~\cite{ma2025} discusses the potential of photocatalytic systems to deliver atmospheric pollutant reductions outside the laboratory in cities, for species including NO$_\text{x}$, VOCs and O$_3$.} 
\updateii{We apply our methodology to O$_3$ in Section S6 and discuss another relevant real-world example of catalytic pollutant removal.}
Building on this evidence, we quantify global removal potential from pollutant fluxes, including high-GWP gases such as CH$_4$, thereby linking air-quality benefits with climate-relevant mitigation. 


\updateii{In forced internal environments such as HVAC systems, additional removal technologies can increase pressure drop and therefore energy consumption. 
However, in this study we focus on integrating removal technologies into existing infrastructure.
For example, integrating catalysts into HVAC duct walls or filters is anticipated to have minimal impact on pressure drop (and could be accommodated through filter design or grade adjustments), whereas introducing additional flow-through removal structures (e.g., supplementary filters or catalytic monoliths) could lead to increases. 
In Europe, HVAC systems consume approximately 313 TWh/y of electricity~\cite{knight2012assessing}; scaled to global consumption, with Europe representing approximately 10\% of global electricity demand, this implies global HVAC electricity consumption of approximately 3100 TWh/y, equivalent to approximately 0.6 GtCO$_2$/y at a grid carbon intensity of 0.2 kgCO$_2$/kWh~\cite{eea2024electricityco2}. 
Only a fraction of this energy is associated with air movement (fan power), which is typically on the order of 20\% of total HVAC electricity consumption. 
If pollutant removal systems increased HVAC pressure drop by approximately 10\%, fan electricity consumption would increase proportionally (assuming constant volumetric flow rate), corresponding to an additional 62 TWh/y or 0.012 GtCO$_2$/y globally. 
For comparison, the CH$_4$ removal potentials estimated here for HVAC filters correspond to approximately $O(10^{-4})$ to $O(10^{0})$ GtCO$_2$e/y (20-year GWP). 
This comparison suggests that pressure-drop penalties could offset the climate benefit of additional removal technologies at low removal efficiencies; however, higher removal rates could remain climate-positive.}
However, integrating surface-based pollutant removal technologies into HVAC systems could build upon their existing capabilities for PM and VOC removal~\cite{liu2020progress, tomlinson2026modelling}, enhancing overall cost-effectiveness and scalability.
A similar integration of technologies was suggested during the COVID-19 pandemic, when HVAC systems were adapted to mitigate airborne viruses, using technologies such as UV-C light and antimicrobial coatings~\cite{thornton2022impact, watson2022efficacy}. 
Furthermore, HVAC filters are regularly replaced, potentially making the maintenance of any surface-based removal technologies straightforward and allowing their performance to be sustained over time.
\updateii{A cost- and climate break-even analysis of these additional removal technologies in HVAC systems is left for future work.}

\update{For context, using a laboratory removal efficiency we estimate a median potential removal rate of CO$_2$ with sorbents using all city surfaces (2 GtCO$_2$/y), corresponding to around 5\% of global anthropogenic CO$_2$ emissions (40 Gt/y)~\cite{friedlingstein2022global}. 
For CH$_4$, the median potential removal rate with catalysis on all city surfaces ($2\times10^{-2}$ GtCH$_4$/y) is around 4\% of the estimated total natural atmospheric CH$_4$ sink of approximately 0.5 GtCH$_4$/y~\cite{dlugokencky2011global}, and about 3–4\% of global CH$_4$ emission rates of around $0.55$ GtCH$_4$/y~\cite{friedlingstein2022global}. 
For NO$_\text{x}$, the median potential removal rate with catalysis on all city surfaces ($2\times10^{-4}$ GtNO$_\text{x}$/y) corresponds to around 0.2\% of NO$_\text{x}$ emission rates of around $0.12$ GtNO$_\text{x}$/y~\cite{mcduffie2020global}.
\updateii{However, the removal potentials estimated here with laboratory removal efficiencies represent transport-limited upper bounds rather than demonstrated technological performance. 
Achieving real-world removal would require catalytic systems capable of sustained operation under atmospheric conditions, together with deployment across very large surface areas. 
Current catalytic technologies typically exhibit lower reaction rates under atmospheric conditions and may require significant energy input~\cite{abernethy2023assessing, tsopelakou2024exploring, Russell2021Review}.
Such removal rates are therefore not currently achievable at scale, and would require substantial technological advances before they could be realised in practice.} 
}

\updateii{If potential removal rates approach the magnitude of natural sinks or anthropogenic emissions, atmospheric concentrations would decrease, reducing concentration-driven flow rates to surfaces. 
While a simple well-mixed box model provides a first-order estimate of concentration evolution, spatial heterogeneity, boundary-layer dynamics and global circulation need to be integrated into atmospheric transport models. 
Future work will couple surface-based removal into pollutant transport and climate models to evaluate regional transport limitations, feedbacks with natural sinks and atmospheric responses.} 
\updateii{
A full assessment must account for regeneration energy, replacement frequency, energy consumption and performance degradation. 
These factors are important for dilute pollutants where large air volumes must be processed, and will determine whether a given technology achieves climate break-even. 
Future work must integrate these engineering inputs with more specific system-level modelling to determine the cost and climate-benefit for the promising environments identified here for atmospheric pollutant removal.} 
\updateii{
Multiplying the atmospheric flow rate to surfaces by literature removal efficiencies should therefore be interpreted as a transport-limited upper-bound estimate of potential pollutant removal. 
Application-specific removal efficiencies measured under characteristic flow velocities, pollutant concentrations, humidity and contact times are not currently available across the range of pollutants and applications considered here; characterising these values represents an important direction for future work. 
}


\update{Scaling surface-based atmospheric pollutant removal technologies presents several additional challenges that must be addressed in future research, including material degradation, surface contamination and infrastructure integration. 
Sorption systems generally maintain predictable CO$_2$ removal efficiencies that decline as sorbent materials become saturated~\cite{soeherman2023overcoming}, whereas the catalytic removal of CH$_4$ is more variable~\cite{tsopelakou2024exploring, pennacchio2024physical}, and can fluctuate under varying conditions (e.g., relative humidity, light intensity). 
Advances in materials and reaction mechanisms can enhance these removal efficiencies and broaden the applicability of catalysis. 
However, large-scale deployment of catalysts may result in undesirable reactions, such as the partial oxidation of CH$_4$ yielding formaldehyde~\cite{pitchai1986}, or the reduction of NO$_\text{x}$ producing N$_2$O~\cite{kamasamudram2012}. 
The distribution of such products can vary depending on catalyst composition, morphology, humidity and pollutant mixtures. 
Mitigation measures include tailoring catalysts to favour complete oxidation of CH$_4$ to CO$_2$ and H$_2$O while suppressing partial oxidation pathways~\cite{pitchai1986}, immobilising catalysts on durable supports to minimise particle release~\cite{zhao2006} and operating under controlled temperature–humidity conditions to reduce N$_2$O formation from NO$_\text{x}$ reduction~\cite{kamasamudram2012}. 
In HVAC systems, catalytic removal could be combined with downstream filtration to capture any harmful intermediates before air is recirculated. 
Additional risks, such as catalyst poisoning by sulphur or siloxane species~\cite{zhou2025}, can be mitigated via upstream pre-filtration and periodic regeneration of catalysts. 
Incorporating these safeguards into deployment strategies may enable the benefits of large-scale catalysis to be realised while protecting environmental and public health. 
}

These findings highlight the potential for integrating surface-based atmospheric pollutant removal technologies into urban and industrial systems to support climate action, while noting that technical advances are required to realise this potential.

\vspace{-0.5cm}
\begin{acknowledgement}

We acknowledge Grantham Foundation for supporting this research.

\end{acknowledgement}

\vspace{-0.5cm}
\begin{suppinfo}

Additional supplementary Figures S1--S2 and Tables S1--S5 are provided in the SI. 

\end{suppinfo}


\vspace{-0.5cm}
\bibliography{achemso-demo}

\end{document}


\setlength{\baselineskip}{.8\baselineskip} 









\section{Ambient concentrations and diffusivities of pollutants}

Understanding the fluxes and flow rates of greenhouse gases (GHGs) and atmospheric pollutants across various environments requires knowledge of their ambient concentrations ($c$), as well as their diffusivities ($D$).
Table~\ref{table:concentrations_diffusivities} presents ranges of typical ambient concentrations and diffusivities of carbon dioxide (CO$_2$), methane (CH$_4$), nitrous oxide (N$_2$O), particulate matter (PM$_{2.5}$), sulphur dioxide (SO$_2$) and nitrogen oxides (NO$_\text{x}$) in the atmosphere.
We have included ranges as atmospheric concentrations can vary inside natural and mechanical environments due to emissions, removal and dispersion~\cite{britter2003flow}. 
We use these six examples to display our results, however, the methods used are general and one would only need to change $c$ and $D$ to model a different atmospheric pollutant.

\section{Estimating parameters for natural and mechanical flows}

This section describes the approach used to collect and estimate the parameter ranges in Tables~\ref{tab:1a}--\ref{tab:1c}.
We provide minimum, mean and maximum values for the length, surface area, velocity, total number and produced number for each example of natural (Section~\ref{subsec:n}), internal (Section~\ref{subsec:fi}) and external (Section~\ref{subsec:fe}) advective flow.
Additionally, we include estimates for heat flux and temperature difference in Section~\ref{subsec:n}; power, specific fan power, cubic feet per minute (CFM) per person, occupancy, CFM per square meter, building floor area, air changes per hour (ACH) and building volume in Section~\ref{subsec:fi}; and drag force and drag coefficient in Section~\ref{subsec:fe}.

\update{Parameter values were primarily derived from literature and measurement datasets. 
Where sufficient data were available (e.g., city velocity distributions, Sec.~\ref{subsubsec:cities}), we directly estimated minimum, mean and maximum values. 
For parameters with limited data, representative literature values were selected for the mean, and a conservative range, typically spanning an order of magnitude, was applied to capture uncertainty. 
These triangular distributions were substituted into the atmospheric pollutant flux and flow rate equations (Sec.~\ref{sec:methods}). 
The resulting distributions of outcomes are summarised in Figs. 1--3 of the main text, with mean values and box-and-whisker plots used to illustrate variability across methods. 
}

\subsection{Natural flows} \label{subsec:n}

\subsubsection{Cities} \label{subsubsec:cities}

The minimum and maximum lengths of cities are assumed to vary between $7 \times 10^3$ m and $5 \times 10^4$ m. 
In Bettencourt \textit{et al.}~\cite{bettencourt2016urban}, the minimum length is reported as $7 \times 10^3$ m, with a corresponding minimum area of $(7 \times 10^3)^2 = 5 \times 10^7$ m$^2$ for Las Palmas, and a maximum length of $5 \times 10^4$ m with an area of $3 \times 10^9$ m$^2$ for London.
Wind speeds in cities, influenced by various local factors~\cite{britter2003flow}, typically range from 1 to 15 m/s.
Measurements by Holzworth \textit{et al.}~\cite{holzworth1972mixing} in several U.S. locations indicate a minimum wind speed of 1 m/s and a maximum of 13 m/s.
The global number of cities is estimated to be between $1 \times 10^3$ and $1 \times 10^4$, depending on definitions based on size or population.
In Angel \textit{et al.}~\cite{angel2011dimensions}, the number of cities worldwide is estimated at 3646. %
They further predict that urban land cover will increase from $3 \times 10^5$ km$^2$ to $1.2 \times 10^6$ km$^2$ over 50 years, suggesting that 60 to 600 new cities are produced annually using Bettencourt \textit{et al.}~\cite{bettencourt2016urban}.
We assume the heat flux in cities ranges from 10 W/m$^2$ to 600 W/m$^2$, supported by evidence in Allen \textit{et al.}~\cite{allen2011global}, where average values range from 10 W/m$^2$ in smaller cities to 600 W/m$^2$ in New York.
Temperature differences between city centres and surrounding areas can range from 1$^\circ$C to 10$^\circ$C.
In Erell \textit{et al.}~\cite{erell2007intra}, temperature differences of this magnitude were observed between urban street canyons and suburban areas in Adelaide. 
The deposition velocity varies from $0.001$ m/s to $0.01$ m/s and the friction velocity varies from $0.1$ m/s to 1 m/s in Giardina \textit{et al.}~\cite{giardina2019atmospheric}.

\renewcommand{\arraystretch}{1.4}
\begin{table*}[t!]
    \centering
    \caption{
     Assumed ambient concentration ranges (mol/m$^3$) and molecular diffusivities (m$^2$/s) of CO$_2$, CH$_4$, N$_2$O, PM$_{2.5}$, SO$_2$ and NO$_\text{x}$, used for modelling atmospheric pollutant behaviour~\cite{friedlingstein2022global, dlugokencky2011global, tian2023global, world2021global, manisalidis2020environmental, boningari2016impact}. %
    } 
    \vspace{0.25cm}
    \resizebox{.45\textwidth}{!}{
    \begin{tabular}{@{}lcc@{}}
    \toprule
    Species & Ambient concentration (mol/m$^3$) & Diffusivity (m$^2$/s) \\ 
    \midrule
    CO$_2$  & $[2\times10^{-2}, \, 4\times10^{-2}]$  & $2 \times 10^{-5}$ \\ 
    CH$_4$  & $[6\times10^{-5}, \, 1\times10^{-4}]$  & $2 \times 10^{-5}$ \\ 
    N$_2$O  & $[8\times10^{-6}, \, 2\times10^{-5}]$  & $2 \times 10^{-5}$ \\ 
    PM$_{2.5}$   & $[1\times10^{-7}, \, 1\times10^{-6}]$  & $1 \times 10^{-8}$ \\ 
    SO$_2$  & $[2\times10^{-7}, \, 5\times10^{-6}]$  & $1 \times 10^{-5}$ \\ 
    NO$_\text{x}$ & $[2\times10^{-6}, \, 8\times10^{-6}]$ & $2 \times 10^{-5}$ \\ 
    \bottomrule
    \end{tabular}}
    \label{table:concentrations_diffusivities}
\end{table*}

Building lengths are assumed to range from 6 to 100 m, with surface areas varying from 40 to $1 \times 10^4$ m$^2$.
In London, Steadman \textit{et al.}~\cite{steadman2009wall} reports building surface areas between 40 m$^2$ for smaller buildings and 10000 m$^2$ for larger ones.
Assuming walls are approximately square (surface area $\sim$ length$^2$), this corresponds to building lengths from 6 m for small residential units to 100 m for large commercial structures.
The current number of buildings worldwide is estimated to range from $3 \times 10^7$ to $1 \times 10^8$, with 2 to 8 million new buildings produced annually.
These estimates align with city building density assumptions of 10 to 40 buildings per hectare~\cite{davies2008city}, resulting in a lower bound of $3 \times 10^7$ and an upper bound of $1 \times 10^8$.
Based on urban expansion rates from Angel \textit{et al.}~\cite{angel2011dimensions}, this yields 2 to 8 million new buildings annually.

\renewcommand{\arraystretch}{1.4}
\begin{table*}[t!]
    \centering
    \caption{
    The minimum, mean and maximum parameter values for the length (m), surface area (m$^2$), velocity (m/s), total number (--), produced number (--), heat flux (\si{\watt/\metre\squared}), surface area (\si{\metre\squared}), temperature difference (\si{\kelvin}) and power (\si{\watt}) are shown. 
    We consider natural flows over cities~\cite{bettencourt2016urban, holzworth1972mixing, angel2011dimensions, allen2011global, erell2007intra}, buildings~\cite{steadman2009wall, holzworth1972mixing, davies2008city, angel2011dimensions}, solar farms~\cite{woolway2024decarbonization, Jones2014Solar, holzworth1972mixing, hossain2024solar, glick2020influence, yang2017study} and solar panels~\cite{saad2020potential, holzworth1972mixing, woolway2024decarbonization, Jones2014Solar}.
    } 
    \vspace{0.5cm}
    \resizebox{.9\textwidth}{!}{
    \begin{tabular}{@{}lcccccccccc@{}}
    \toprule
    \text{Environment} & \text{Parameter} & \text{Ref.} & \text{Min.} & \text{Mean} & \text{Max.} & \text{Parameter} & \text{Ref.} & \text{Min.} & \text{Mean} & \text{Max.} \\
    \midrule
    \multirow{4}{*}{Cities} & Length (\si{\metre}) & \cite{bettencourt2016urban} & $7 \times 10^{3}$ & $1 \times 10^{4}$ & $5 \times 10^{4}$ & Total number (\si{-}) & \cite{angel2011dimensions} & $1 \times 10^{3}$ & $5 \times 10^{3}$ & $1 \times 10^{4}$ \\
    & Surface area (\si{\metre\squared}) & \cite{bettencourt2016urban} & $5 \times 10^{7}$ & $1 \times 10^{8}$ & $3 \times 10^{9}$ & Produced number (\si{-}) & \cite{angel2011dimensions} & $100$ & $200$ & $400$ \\
    & Velocity (\si{\metre/\second}) & \cite{holzworth1972mixing} & $1$ & $4$ & $15$ & Heat flux (\si{\watt/\metre\squared}) & \cite{allen2011global} & $1 \times 10^{1}$ & $1 \times 10^{2}$ & $6 \times 10^{2}$ \\
    & Temperature difference (\si{\celsius}) & \cite{erell2007intra} & $1$ & $3$ & $10$ & & & & & \\
    \midrule
    \multirow{3}{*}{Buildings} & Length (\si{\metre}) & \cite{steadman2009wall} & $6$ & $30$ & $100$ & Total number (\si{-}) & \cite{davies2008city} & $1 \times 10^{7}$ & $5 \times 10^{7}$ & $1 \times 10^{8}$ \\
    & Surface area (\si{\metre\squared}) & \cite{steadman2009wall} & $40$ & $1 \times 10^3$ & $1 \times 10^4$ & Produced number (\si{-}/y) & \cite{angel2011dimensions} & $2 \times 10^6$ & $4 \times 10^6$ & $8 \times 10^6$ \\
    & Velocity (\si{\metre/\second}) & \cite{holzworth1972mixing} & $1$ & $4$ & $15$ & & & & & \\
    \midrule
    \multirow{4}{*}{Solar farms} & Length (\si{\metre}) & \cite{woolway2024decarbonization} & $300$ & $750$ & $7 \times 10^{3}$ & Total number (\si{-}) & \cite{hossain2024solar} & $2 \times 10^{3}$ & $1 \times 10^{4}$ & $5 \times 10^{4}$ \\
    & Surface area (\si{\metre\squared}) & \cite{woolway2024decarbonization} & $1 \times 10^5$ & $5 \times 10^5$ & $5 \times 10^7$ & Produced number (\si{-}) & \cite{hossain2024solar} & $500$ & $2 \times 10^{3}$ & $5 \times 10^3$ \\
    & Velocity (\si{\metre/\second}) & \cite{holzworth1972mixing} & $1$ & $4$ & $15$ & Heat flux (\si{\watt/\metre\squared}) & \cite{glick2020influence, yang2017study} & $1$ & $5$ & $15$ \\
    & Temperature difference (\si{\celsius}) & \cite{glick2020influence, yang2017study} & $0.1$ & $2$ & $5$ & & & & & \\
    \midrule
    \multirow{3}{*}{Solar panels} & Length (\si{\metre}) & \cite{saad2020potential} & $1$ & $4$ & $10$ & Total number (\si{-}) & \cite{hossain2024solar}  & $1 \times 10^{6}$ & $5 \times 10^6$ & $7 \times 10^{7}$ \\
    & Surface area (\si{\metre\squared}) & \cite{woolway2024decarbonization} & $1$ & $16$ & $100$ & Produced number (\si{-}/y) & \cite{hossain2024solar} & $1 \times 10^{7}$ & $5 \times 10^{7}$ & $1 \times 10^{8}$ \\
    & Velocity (\si{\metre/\second}) & \cite{holzworth1972mixing} & $1$ & $4$ & $15$ & & & & & \\
    \bottomrule
    \end{tabular}}
    \label{tab:1a}
\end{table*}

\subsubsection{Solar farms} \label{subsubsec:solar_farms}

We assume that small solar farms typically span around 300 m in length and cover an area of $1 \times 10^{5}$ m$^2$, while large solar farms can extend up to 7000 m in length, covering an area of $5 \times 10^{7}$ m$^2$.
These ranges are supported by Woolway \textit{et al.} and Jones \textit{et al.}~\cite{woolway2024decarbonization, Jones2014Solar}, which describe small installations covering $1 \times 10^{5}$ m$^2$ and the largest solar farms reaching up to $5 \times 10^{7}$ m$^2$.
Globally, the number of solar farms is estimated to be between $1 \times 10^3$ and $2 \times 10^4$, with 500 to 5000 new farms being established annually.
According to Hossain \textit{et al.}\cite{hossain2024solar}, the global installed capacity of solar farms reached 1400 GW in 2023, up from 1100 GW in 2022.
Assuming an average capacity of 100 MW per farm, this translates to approximately 20000 solar farms worldwide and about 3000 new farms annually.
We assume wind speeds over solar farms are consistent with those reported for cities in Holzworth \textit{et al.}~\cite{holzworth1972mixing}.
We estimate the heat flux across solar farms to range from 1 to 30 W/m$^2$, with temperature differences varying between 1$^\circ$C and 8$^\circ$C.
In Glick \textit{et al.}~\cite{glick2020influence}, a heat flux of approximately 1 W/m$^2$ is reported, while Yang \textit{et al.}~\cite{yang2017study} suggests a net radiation of around 100 W/m$^2$, implying a heat flux of up to 30 W/m$^2$.
Both studies report temperature differences of 1$^\circ$C to 8$^\circ$C.

Solar panel dimensions are assumed to range from 1 m for small panels to 10 m for large installations.
In Saad \textit{et al.}\cite{saad2020potential}, panel dimensions are generally around 2 m; however, we also account for the potential stacking of panels in large farms.
Consequently, panel surface areas are expected to vary from 1 m$^2$ to 100 m$^2$. 
The total global number of solar panels is estimated to be between $5 \times 10^7$ and $5 \times 10^8$, with new installations ranging from $1 \times 10^7$ to $4 \times 10^8$.
Based on the panel size and farm capacities in Woolway \textit{et al.} and Jones \textit{et al.}~\cite{woolway2024decarbonization, Jones2014Solar}, we estimate that the average solar farm has about $5 \times 10^4$ panels, giving a total of $5 \times 10^8$ panels globally, with about $1 \times 10^8$ new panels produced each year.

\subsection{Internal flows} \label{subsec:fi}

\subsubsection{HVAC systems} \label{subsubsec:hvac}

\renewcommand{\arraystretch}{1.4}
\begin{table*}[t!]
    \centering
    \caption{
    The minimum, mean, and maximum parameter values for the diameter (m), surface area (m$^2$), velocity (m/s), total number (--), produced number (--), specific fan power (\si{\watt/(\metre\cubed/\second)}), CFM per person (\text{ft}$^3$/\text{min}), number of people (--), CFM per unit area (\text{ft}$^3$/\text{m}$^2$), building floor area (\si{\metre\squared}), air changes per hour (1/h), building volume (\si{\metre\cubed}), length (\si{\metre}), pleat number (--) and efficiency (\%) for various systems. 
    We consider HVAC systems~\cite{Fisk2000Duct, fisk2004evaluation, knight2012assessing, merzkirch2016field, eto1990hvac, herrmann2022global, persily2004analysis, memarzadeh2012role}, combustion systems~\cite{Fisk2000Duct, hale2014effect, zhang2007development, carbonbrief2020coal, knight2012assessing, merzkirch2016field}, DAC systems~\cite{Fisk2000Duct, tegeler2023novel, IEA2023_2, knight2012assessing, merzkirch2016field} and HVAC filters~\cite{schroth1996new, al2009effect, lowther2023factors}.
    } 
    \vspace{0.5cm}
    \resizebox{\textwidth}{!}{
    \begin{tabular}{@{}lcccccccccc@{}}
    \toprule
    \text{System} & \text{Parameter} & \text{Ref.} & \text{Min.} & \text{Mean} & \text{Max.} & \text{Parameter} & \text{Ref.} & \text{Min.} & \text{Mean} & \text{Max.} \\
    \midrule
    \multirow{7}{*}{HVAC} 
    & Diameter (\si{\metre}) & \cite{Fisk2000Duct} & $0.2$ & $0.7$ & $2.0$ & CFM per person (ft$^3$/min) & \cite{eto1990hvac} & $20$ & $35$ & $50$ \\
    & Surface area (\si{\metre\squared}) & \cite{Fisk2000Duct} & $300$ & $1\times10^3$ & $2\times10^3$ & People (\si{-}) & \cite{herrmann2022global} & $2\times10^{9}$ & $4\times10^{9}$ & $5\times10^{9}$ \\
    & Velocity (\si{\metre/\second}) & \cite{fisk2004evaluation} & $2$ & $3.5$ & $5$ & CFM per m$^2$ (ft$^3$/m$^2$) & \cite{persily2004analysis} & $0.1$ & $1.5$ & $3$ \\
    & Total number (\si{-}) & \cite{davies2008city} & $2 \times 10^{8}$ & $5 \times 10^{8}$ & $1 \times 10^{9}$ & Floor area (\si{\metre\squared}) & \cite{persily2004analysis} & $1\times10^{11}$ & $4\times10^{11}$ & $1\times10^{12}$ \\
    & Produced number (\si{-}) & \cite{angel2011dimensions} & $1 \times 10^{7}$ & $2 \times 10^{7}$ & $3 \times 10^{7}$ & Air changes per hour (\si{1/h}) & \cite{memarzadeh2012role} & $0.5$ & $3$ & $5$ \\
    & Power (\si{\watt}) & \cite{knight2012assessing} & $1 \times 10^{8}$ & $1 \times 10^{10}$ & $1 \times 10^{11}$ & Volume (\si{\metre\cubed}) & \cite{persily2004analysis} & $6\times10^{11}$ & $1\times10^{12}$ & $2\times10^{12}$ \\
    & Specific fan power (\si{\watt/(\metre\cubed/\second)}) & \cite{merzkirch2016field} & $0.2$ & $1.5$ & $2.5$ & & & & & \\
    \midrule
    \multirow{4}{*}{Combustion} 
    & Diameter (\si{\metre}) & \cite{Fisk2000Duct} & $0.2$ & $0.7$ & $2.0$ & Produced number (\si{-}) & \cite{carbonbrief2020coal} & $500$ & $750$ & $1\times10^3$ \\
    & Surface area (\si{\metre\squared}) & \cite{hale2014effect} & $50$ & $275$ & $500$ & Power (\si{\watt}) & \cite{carbonbrief2020coal, knight2012assessing} & $1 \times 10^{8}$ & $2 \times 10^{10}$ & $1 \times 10^{13}$ \\
    & Velocity (\si{\metre/\second}) & \cite{zhang2007development} & $10$ & $20$ & $30$ & Specific fan power (\si{\watt/(\metre\cubed/\second)}) & \cite{merzkirch2016field} & $0.2$ & $1.5$ & $2.5$\\
    & Total number (\si{-}) & \cite{carbonbrief2020coal} & $1 \times 10^{4}$ & $2 \times 10^{4}$ & $5 \times 10^{4}$ & & & & & \\
    \midrule
    \multirow{4}{*}{DAC} 
    & Diameter (\si{\metre}) & \cite{Fisk2000Duct} & $0.2$ & $0.7$ & $2.0$ & Produced number (\si{-}) & \cite{IEA2023_2} & $10$ & $50$ & $150$ \\
    & Surface area (\si{\metre\squared}) & \cite{mcqueen2021review} & $100$ & $300$ & $500$ & Power (\si{\watt}) & \cite{IEA2023_2, knight2012assessing} & $1 \times 10^{7}$ & $3 \times 10^{8}$ & $5 \times 10^{9}$ \\
    & Velocity (\si{\metre/\second}) & \cite{tegeler2023novel} & $1$ & $2$ & $3$ & Specific fan power (\si{\watt/(\metre\cubed/\second)}) & \cite{merzkirch2016field} & $0.2$ & $1.5$ & $2.5$\\
    & Total number (\si{-}) & \cite{IEA2023_2} & $50$ & $300$ & $1\times10^3$ & & & & & \\
    \midrule
    \multirow{3}{*}{Filters} 
    & Length (\si{\metre}) & \cite{schroth1996new} & $0.2$ & $0.7$ & $2.0$ & Efficiency (\%) & \cite{schroth1996new} & $97.0$ & $99.0$ & $100.0$ \\
    & Surface area (\si{\metre^2}) & \cite{al2009effect} & $0.2$ & $20$ & $200$ & Total number (\si{-}) & \cite{lowther2023factors} & $5 \times 10^{8}$ & $1 \times 10^{9}$ & $2 \times 10^{9}$ \\
    & Velocity (\si{\metre/\second}) & \cite{tegeler2023novel} & $0.09$ & $0.1$ & $0.4$ & Produced number (\si{-}/y) & \cite{lowther2023factors} & $1 \times 10^{8}$ & $3 \times 10^{8}$ & $4 \times 10^{8}$ \\ 
    \bottomrule
    \end{tabular}}
    \label{tab:1b}
\end{table*}

Average duct diameters are assumed to range from approximately 0.2 to 2 m.
In large commercial buildings, Fisk \textit{et al.}\cite{Fisk2000Duct} reports that smaller ducts have cross-sectional dimensions of 15--30 cm, while larger ducts can reach several meters.
We estimate duct surface areas to vary from 300 to 2000 m$^2$. 
In Fisk \textit{et al.}\cite{Fisk2000Duct}, the authors examine HVAC systems in both small and large buildings, with duct surface areas ranging from 220–710 m$^2$ in smaller buildings and 1400–3400 m$^2$ in larger ones.
Air velocity in HVAC ducts is typically estimated to fall between 2 and 5 m/s. %
In Fisk \textit{et al.}\cite{fisk2004evaluation}, maximum air velocities measured in their systems ranged from 0.8 to 3 m/s, consistent with our assumed range.
Our analysis uses the global building counts and new construction rates discussed in Section~\ref{subsubsec:cities}.
The power consumption of HVAC systems is assumed to range from 100 MW to 100 GW. %
According to Knight \textit{et al.}\cite{knight2012assessing}, 313 TWh are consumed annually in Europe alone, which translates to approximately $313 \times 10^{12} / 3600 = 9 \times 10^{10}$ W, fitting within our estimated range.
The deposition velocity varies from $10^{-6}$ m/s to $1$ m/s in Sippola \textit{et al.}~\cite{sippola2002particle}.

We assume the specific fan power varies from 0.5 to 2.5 W/(m$^3$/s), aligning with field measurements in Merzkirch \textit{et al.}\cite{merzkirch2016field}, where specific fan power ranged from 0.23 to 1.9 W/(m$^3$/s).
Ventilation capacity per person is assumed to range from 20 to 50 cubic feet per minute (CFM).
A reported value of 35 CFM per person for medium-sized U.S. offices in Eto \textit{et al.}\cite{eto1990hvac} is within this range.
The global working population is estimated at 2 to 5 billion people, with a midpoint of 4 billion, or approximately 50\% of the global population, aligning with estimates from Herrmann \textit{et al.}\cite{herrmann2022global}.
For ventilation per floor area, we assume values from 0.5 to 4 CFM per m$^2$, consistent with measurements in Persily \textit{et al.}\cite{persily2004analysis}, where the CFM per m$^2$ varied from 0.1 to 3.
We estimate the global floor area at $1 \times 10^{11}$ to $1 \times 10^{12}$ m$^2$, validated by typical floor areas of 100 to 3040 m$^2$ in Persily \textit{et al.}\cite{persily2004analysis} multiplied by the estimated building count ($5 \times 10^8 \times 3 \times 10^3 = 1 \times 10^{12}$ m$^2$).
Air changes per hour (ACH) are assumed to range from 0.5 to 5.
While Memarzadeh \textit{et al.}\cite{memarzadeh2012role} reports ACH values of 4 to 12, the higher end generally applies to hospital environments. 
For general building volumes, we estimate a range from $6 \times 10^{11}$ to $2 \times 10^{12}$ m$^3$.
This estimate is validated by multiplying the global floor area from Persily \textit{et al.}\cite{persily2004analysis} by the building count and an assumed ceiling height of 2 m, yielding $2 \times 10^{12}$ m$^3$.

Note that we could further divide these HVAC systems into buildings, clean rooms and data centres, which perturbs the representative duct dimensions, velocities and total/produced numbers that we would have to find in the literature or using industry standards like SFP, CFM or ACH.
However, this does not introduce any new physics to these internal flows and so is neglected from our analysis.

\subsubsection{Combustion systems}

Fossil fuel combustion chambers require robust ventilation systems to maintain the high airflow essential for combustion. 
We assume the duct diameters in these systems align with the range specified in Section~\ref{subsubsec:hvac}.
The surface area of these systems is estimated to range between 50 and 800 m$^2$, supported by Hale \textit{et al.}\cite{hale2014effect}, which reports power plant sizes ranging from 200 to 800 m. Assuming a typical duct diameter of 1 m, this would imply similar duct lengths within our assumed range.
Higher airflow rates that standard HVAC systems, estimated between 10 and 30 m/s, are necessary to ensure efficient fuel combustion.
As noted in Zhang \textit{et al.}\cite{zhang2007development}, an airflow velocity of 20 m/s is common, sitting centrally within this range.
Globally, we estimate the number of power plants ranges from $1 \times 10^4$ to $5 \times 10^4$, with annual additions of 100 to 1000 new plants.
The operational estimate of \(1 \times 10^4\) plants given by Carbon Brief\cite{carbonbrief2020coal} supports the lower end of this spectrum.
Based on operational power data from the past decade, which spans $1.6 \times 10^6$ to $1.8 \times 10^6$ MW, we estimate around 120 new power plants are added each year. 
Using these figures alongside the energy and specific fan power metrics from Section~\ref{subsubsec:hvac}, we estimate that the power required for ventilation in these systems ranges from $5 \times 10^{3}$ to $3 \times 10^{4}$ W.
Specific fan power is assumed to vary between 0.5 and 2.5 W/(m$^3$/s), consistent with similar industrial applications.

\subsubsection{Direct air capture (DAC) systems}

We maintain the mean diameter of DAC ventilation systems within the range of 0.2–2 m, as specified in Section~\ref{subsubsec:hvac}.
The estimated surface area of DAC systems ranges from 100 to 500 m$^2$, consistent with dimensions observed in large DAC plants, which measure around 100 m in length~\cite{mcqueen2021review}.
Velocity estimates, based on airflow needed to capture CO$_2$ effectively from ambient air, generally fall between 1 and 6 m/s.
This range is confirmed in Tegeler \textit{et al.}~\cite{tegeler2023novel}, where the mean velocity is reported as 3 m/s.
Current DAC installations are estimated at 20 to 40 units, with new installations projected to increase by 10 to 30 units annually~\cite{IEA2023_2}.
Utilising the total number of plants and the energy and specific fan power values discussed in Section~\ref{subsubsec:hvac}, we estimate energy requirements for DAC systems to range from 400 to 800 W.
This energy range accommodates variations in system size and efficiency, with specific fan power requirements between 1 and 2.5 W/(m$^3$/s) for effective operation. 

\subsubsection{HVAC filters}

The length of HVAC filter sheet is generally between 0.01 and 0.3 meters.
In Schroth \textit{et al.}~\cite{schroth1996new}, for example, a HVAC filter sheet used in clean rooms is 0.28 meters long.
We assume the number of pleats in a HVAC filter ranges from 100 to 300.
According to Al \textit{et al.}~\cite{al2009effect}, the number of pleats per 100 mm ranges from 28 to 34; with a filter width of 592 mm, this translates to 166 to 201 pleats, which aligns with our assumed range.
This means that the surface area of the HVAC filter sheet (length $\times$ width $\times$ number of pleats) ranges from 0.2 to 200 m$^2$.
Note that, e.g., when evaluating flow rates using the surface area, 0.2 m$^2$, the corresponding velocity must be used, 0.4 m/s, as the flow rate through the filter must remain constant, 0.08 m$^3$/s.
HVAC filter efficiency is estimated between 95\% and 100\%, consistent with efficiencies reported for filter classes EU11 to EU14 in Schroth \textit{et al.}~\cite{schroth1996new}.
We estimate the total number of HVAC filters in use worldwide to be between \(5 \times 10^8\) and \(2 \times 10^9\), with annual production levels around \(8 \times 10^7\) to \(4 \times 10^9\).
This estimation is consistent with the assumption that the number of filters aligns closely with the global building count, while the higher annual production reflects a growing demand for improved air quality.
These estimates also align with data from Lowther \textit{et al.}~\cite{lowther2023factors}, where the number of HVAC units sold in China in 2017 was 982 million, and the annual production was approximately 145 million.

\subsection{External flows} \label{subsec:fe}

\subsubsection{Aeroplanes}

\renewcommand{\arraystretch}{1.4}
\begin{table*}[t!]
    \centering
    \caption{
    The minimum, mean and maximum parameter values for length (m), surface area (m$^2$), velocity (m/s), total number (--), produced number (--), drag force (\si{\newton}) and drag coefficient (--).
    We consider flows around aeroplanes~\cite{mcmasters1998advanced, filippone2000data, klower2021quantifying, sadraey2009drag}, trains~\cite{schafer2008relationship, allotta2002train, singh2012planes, mario2020numerical} and automobiles~\cite{wasielewski1981effect, wood2003velocity, sperling2009two, sovran2012aerodynamic}.
    } 
    \vspace{0.5cm}
    \resizebox{.8\textwidth}{!}{
    \begin{tabular}{@{}lcccccccccc@{}}
    \toprule
    \text{Environment} & \text{Parameter} & \text{Ref.} & \text{Min.} & \text{Mean} & \text{Max.} & \text{Parameter} & \text{Ref.} & \text{Min.} & \text{Mean} & \text{Max.} \\
    \midrule
    \multirow{4}{*}{Aeroplanes} & Length (\si{\metre}) & \cite{mcmasters1998advanced} & $20$ & $40$ & $80$ & Total number (\si{-}) & \cite{klower2021quantifying} & $2 \times 10^{4}$ & $3 \times 10^{4}$ & $4 \times 10^{4}$ \\
    & Surface area (\si{\metre\squared}) & \cite{mcmasters1998advanced} & $150$ & $300$ & $1000$ & Produced number (\si{-}) & \cite{klower2021quantifying} & $800$ & $1 \times 10^{3}$ & $2 \times 10^{3}$ \\
    & Velocity (\si{\metre/\second}) & \cite{filippone2000data} & $200$ & $250$ & $300$ & Drag force (\si{\newton}) & \cite{sadraey2009drag} & $1 \times 10^{4}$ & $5 \times 10^{4}$ & $2 \times 10^{5}$ \\
    & Drag coefficient (\si{-}) & \cite{sadraey2009drag} & $0.02$ & $0.03$ & $0.05$ & & & & & \\
    \midrule
    \multirow{4}{*}{Trains} & Length (\si{\metre}) & \cite{schafer2008relationship} & $50$ & $200$ & $400$ & Total number (\si{-}) & \cite{singh2012planes} & $1 \times 10^{5}$ & $2 \times 10^{5}$ & $3 \times 10^{5}$ \\
    & Surface area (\si{\metre\squared}) & \cite{schafer2008relationship} & $200$ & $800$ & $2 \times 10^{3}$ & Produced number (\si{-}) & \cite{singh2012planes} & $3 \times 10^{3}$ & $6 \times 10^{3}$ & $1 \times 10^{4}$ \\
    & Velocity (\si{\metre/\second}) & \cite{allotta2002train} & $40$ & $60$ & $100$ & Drag force (\si{\newton}) & \cite{mario2020numerical} & $2 \times 10^{3}$ & $1 \times 10^{4}$ & $3 \times 10^{4}$ \\
    & Drag coefficient (\si{-}) & \cite{mario2020numerical} & $0.2$ & $0.3$ & $0.4$ & & & & & \\
    \midrule
    \multirow{4}{*}{Automobiles} & Length (\si{\metre}) & \cite{wasielewski1981effect} & $3.0$ & $4.5$ & $6.0$ & Total number (\si{-}) & \cite{sperling2009two} & $1 \times 10^{9}$ & $2 \times 10^{9}$ & $3 \times 10^{9}$ \\
    & Surface area (\si{\metre\squared}) & \cite{wasielewski1981effect} & $7$ & $10$ & $18$ & Produced number (\si{-}) & \cite{sperling2009two} & $6 \times 10^{7}$ & $8 \times 10^{7}$ & $1 \times 10^{8}$ \\
    & Velocity (\si{\metre/\second}) & \cite{wood2003velocity} & $30$ & $35$ & $50$ & Drag force (\si{\newton}) & \cite{sovran2012aerodynamic} & $1 \times 10^{2}$ & $3 \times 10^{2}$ & $1 \times 10^{3}$ \\
    & Drag coefficient (\si{-}) & \cite{sovran2012aerodynamic} & $0.2$ & $0.4$ & $0.7$ & & & & & \\
    \bottomrule
    \end{tabular}}
    \label{tab:1c}
\end{table*}

The length of aeroplanes was estimated by considering a range of aircraft models, from the Boeing 737-700 to the Boeing 747-8, with lengths varying between approximately 20 m and 80 m.
Values provided in McMasters \textit{et al.}~\cite{mcmasters1998advanced} (40--71 m) fall within this range.
Using a typical fuselage diameter of 5 m, the estimated body surface area (area $\sim$ $\pi$ $\times$ radius $\times$ length) of aeroplanes ranges from $\pi \times 2.5 \times 20 = 1.5 \times 10^2$ m$^2$ to $\pi \times 2.5 \times 80 = 5 \times 10^2$ m$^2$, which we double to include the wing surface area.
We assume the mean velocity to be between 200 m/s and 300 m/s. 
Subsonic civil/commercial aircraft examined in Filippone \textit{et al.}~\cite{filippone2000data} operate at Mach numbers ranging from 0.6 to 1, supporting these assumptions. 
The estimated total number of aeroplanes worldwide ranges between \(2 \times 10^4\) and \(4 \times 10^4\), with annual production between \(5 \times 10^2\) and \(2 \times 10^3\). %
These values are comparable to the number of flights per day in Europe, reported in Klower \textit{et al.}~\cite{klower2021quantifying} to vary between \(2 \times 10^4\) and \(4 \times 10^4\) over ten years.
Drag forces for aircraft range from \(1 \times 10^4\) N to \(2 \times 10^5\) N, with drag coefficients between 0.01 and 0.05, reflecting the aerodynamic efficiency of various aircraft designs.
In Sadraey \textit{et al.}~\cite{sadraey2009drag}, the drag force and drag coefficient range from \(2 \times 10^4\) to \(2 \times 10^5\) N and 0.02 to 0.04, respectively.

\subsubsection{Trains}

Train lengths were estimated to range from 50 m for short urban light rail trains to 400 m for long-distance freight or high-speed trains.
Assuming an average train car length of 20 m, the mode lengths in Schafer \textit{et al.}~\cite{schafer2008relationship}, ranging from 200 to 400 m, corroborate this estimate.
With a typical train diameter of 4 m (double the standard gauge railway width of 2 m), the surface area is estimated to range from $4 \times 2 \times 50 = 4 \times 10^2$ m$^2$ to $4 \times 2 \times 400 = 3 \times 10^3$ m$^2$.
Train speeds range from 40 m/s for commuter or freight trains to 100 m/s for high-speed rail systems.
In Allotta \textit{et al.}~\cite{allotta2002train}, speeds are estimated to be between 40 and 70 m/s, consistent with our assumptions.
The total number of trains worldwide is estimated to be between \(1 \times 10^5\) and \(3 \times 10^5\), with annual production between \(3 \times 10^3\) and \(1 \times 10^4\).
The number of trains in China is reported to be approximately \(1 \times 10^4\), with annual production around \(3 \times 10^2\), in line with global estimates if China is assumed to represent 10\% of the world’s trains~\cite{singh2012planes}.
We estimate the drag force for trains to range from \(2 \times 10^3\) N to \(3 \times 10^4\) N, with drag coefficients between 0.2 and 0.5, depending on the train design.
In Mario \textit{et al.}~\cite{mario2020numerical}, drag forces of \(1 \times 10^4\) N and drag coefficients of 0.5 were reported, supporting our estimates.

\subsubsection{Automobiles}

Automobile lengths were estimated to range from 3 m for smaller cars to 6 m for larger vehicles such as trucks.
Values reported in Wasielewski \textit{et al.}~\cite{wasielewski1981effect} for medium-length cars (4.57--5.46 m) fall within this range.
The surface area was estimated using an average vehicle width of 2 m, with values ranging from 7 m$^2$ for smaller cars to 18 m$^2$ for larger vehicles with greater height.
Velocity estimates for automobiles were based on typical highway speeds, with a minimum of 10 m/s and a maximum of 40 m/s.
Wood \textit{et al.}\cite{wood2003velocity} provides velocity estimates between 30 and 50 m/s, though the upper limit seems high for average values.
The global number of automobiles is estimated to range between \(1 \times 10^9\) and \(3 \times 10^9\), with annual production between \(6 \times 10^7\) and \(1 \times 10^8\).
Sperling \textit{et al.}\cite{sperling2009two} estimates the global vehicle population approaching \(2 \times 10^9\), with annual growth of about 3\% (\(6 \times 10^7\)).
We assume that 10\% of cars are in use at any given time, contributing to the estimated flow rates in our analysis.
Drag forces for automobiles range from 100 N for small cars to 1000 N for large trucks, with drag coefficients between 0.2 and 0.4, depending on vehicle design.
Sovran \textit{et al.}\cite{sovran2012aerodynamic} reports drag forces and coefficients between 68 and 774 N, and 0.3 to 0.6, respectively, aligning with these estimates.  

\section{Methods} \label{sec:methods}

In this section, we provide a detailed explanation of the calculations used to estimate the flow rate of atmospheric pollutants to the surfaces of various applications, including cities, HVAC systems and transportation.
These calculations are based on scaling theory (Section~\ref{scaling}), empirical relationships (Section~\ref{empirical}) and velocity and atmospheric pollutant profiles (Section~\ref{profiles}).
Additionally, we describe calculations that estimate the flow rate through the boundary layer, incorporating energy measurements, industry standards and drag measurements (Section~\ref{subsec:epandd}).

\subsection{Scaling theory} \label{scaling}

This subsection presents a scaling theory for the pollutant flow rate to a flat surface, considering turbulent boundary layers.
Consider a flat surface exposed to a fluid with a free-stream velocity \( u \) (m/s) and a characteristic length scale \( l \) (m).
The total surface area of the flat surface is \( s \) (m$^2$), where \( n \) denotes the number of flat surfaces (--).
Let \( c \) represent the pollutant concentration in the bulk fluid (mol/m$^3$) and \( D \) be the diffusion coefficient of the pollutant (m$^2$/s).
The pollutant flux \( j_y \) (mol/m$^2$/s) to the flat surface is given by the diffusive flux only, as the advective flux is zero, such that
\begin{equation}
j_y = - D \left( \frac{\partial c}{\partial y} \right)_{y=0},
\end{equation}
where \( \partial c/\partial y \) is the concentration gradient normal to the surface at \( y = 0 \).
Assuming a linear concentration profile in the boundary layer, the gradient at the flat surface can be approximated by
\begin{equation}
\left( \frac{\partial c}{\partial y} \right)_{y=0} \sim \frac{c}{\delta_c}.
\end{equation}
Here, \( \delta_c \) (m) denotes the pollutant boundary layer thickness.
Therefore, the diffusive flux of the pollutant to the flat surface is
\begin{equation}
j_y \sim \frac{D c}{\delta_c}.
\end{equation}
The pollutant boundary layer thickness \( \delta_c \) is related to the momentum boundary layer thickness \( \delta \) through the Schmidt number \( Sc \), defined as \( Sc = \nu/D \), where \( \nu \) is the kinematic viscosity of the fluid (m$^2$/s).
The relationship is given by Frank \textit{et al.}~\cite{frank2017incropera}
\begin{equation}
\delta_c \sim \frac{\delta}{Sc^{1/3}}.
\end{equation}
The total pollutant flow rate \( q_y \) (mol/s) to the flat surface is obtained by multiplying the diffusive flux \( j_y \) by the surface area \( s \)
\begin{equation}
q_y = j_y n s \sim \frac{n s D c}{\delta_c}. 
\end{equation}
For turbulent flow, the turbulent diffusivity \( D + D_e \) can be expressed in terms of the friction velocity \( u^* \) and the von K\'{a}rm\'{a}n constant \( \kappa \) \cite{pope2000}, such that
\begin{equation}
D_e \sim \kappa u^* \delta.  
\end{equation}
Therefore, the pollutant flow rate to the flat surface is
\begin{equation} \label{eq:scaling}
q_y \sim \frac{n s (D + D_e) c }{\delta/{Sc}^{1/3}}.  
\end{equation}
Note that these approximations may not fully capture real-world complexities, particularly for PM, which is influenced by inertia and gravitational settling.  
These species may require different modelling approaches compared to GHGs~\cite{turpin2000measuring}.  

\subsection{Empirical relationships} \label{empirical}

This subsection presents empirical relationships for the boundary layer thickness, drag coefficient, Nusselt number and Sherwood number for a turbulent flow over a flat plate.   
All the empirical relations in this paragraph are taken from Frank \textit{et al.}~\cite{frank2017incropera}.  
The momentum boundary layer thickness is given by 
\begin{equation}
\delta \approx \frac{0.4 l}{Re^{1/5}}.  
\end{equation}
The drag coefficient \( C_d \) for momentum transfer is
\begin{equation}
C_d \approx \frac{0.1}{Re^{1/5}}.  
\end{equation}
The Nusselt number \( Nu \) for heat transfer is
\begin{equation}
Nu  \approx 0.03 Re^{4/5} Pr^{1/3}.  
\end{equation}
The Sherwood number \( Sh \) for mass transfer is
\begin{equation} \label{eq:emp_sh}
Sh \approx 0.03 Re^{4/5} Sc^{1/3}.  
\end{equation}
To relate heat and mass transfer, we use the relationship between the Nusselt and Sherwood numbers, along with the Prandtl and Schmidt numbers, defined as
\begin{equation}
\frac{Nu}{Pr^{1/3}} = \frac{Sh}{Sc^{1/3}}.  
\end{equation}
The mass transfer coefficient \( m_{c} = {Sh} D / l \) can be determined using
\begin{equation} \label{eq:m_c_1}
m_{c} \approx  \frac{0.03 {Re}^{4/5} {Sc}^{1/3} D}{l},
\end{equation}
using \eqref{eq:emp_sh}.  
Therefore, the total pollutant flow rate \( q_y = j_y n s = m_c n s c \) to the flat surface can be calculated using 
\begin{equation} 
\label{eq:empirical}
q_y \approx \frac{0.03 {Re}^{4/5} {Sc}^{1/3} n s D c}{l}.  
\end{equation}
It is important to note that empirical relationships may not fully capture the complexities of real-world systems, especially in cases involving irregular geometries, highly turbulent flows or non-standard flow regimes~\cite{schlichting2016boundary}.  
Under such conditions, deviations from theoretical predictions are expected and adjustments may be necessary to account for these influences.  



\subsection{Fully-developed profiles} \label{profiles}

\subsubsection{Monin–Obukhov similarity theory}

Monin-Obukhov (M-O) similarity theory describes the vertical profiles of wind velocity, temperature and concentrations near the surface.  
Under neutral stability conditions, the velocity profile for turbulent flow near the surface is given by Schlichting \textit{et al.}~\cite{schlichting2016boundary}
\begin{equation}
    u(y) \approx \frac{u^*}{\kappa} \ln\left(\frac{y}{y_0}\right),
\end{equation}
where \(u(y)\) is the wind velocity at height \(y\), \(u^*\) is the friction velocity, \(\kappa\) is the von K\'{a}rm\'{a}n constant and \(y_0\) is the roughness length.   
By the analogy between momentum and mass transfer, the pollutant concentration field \(c(y)\) near the surface can also be assumed to follow a logarithmic profile
\begin{equation} \label{eq:mo_0}
    c(y) \approx c_0 - \frac{c^*}{\kappa} \ln\left(\frac{y}{y_0}\right),
\end{equation}
where \(c_0\) is the reference concentration at a height \(y_0\) and \(c^*\) is the characteristic concentration difference (analogous to the friction velocity).   
The pollutant flux \(j_y = j_y(y)\) at height \(y\) is governed by
\begin{equation} \label{eq:mo_1}
    j_y(y) = vc -(D + D_e) \frac{\partial c}{\partial y}.   
\end{equation}
Using \eqref{eq:mo_1}, the concentration gradient \(\partial c / \partial y\) is
\begin{equation} \label{eq:mo_2}
    \frac{\partial c}{\partial y} \approx \frac{c^*}{\kappa y}.   
\end{equation}
Substituting \eqref{eq:mo_2} into \eqref{eq:mo_1} gives
\begin{equation}
    j_y \approx vc -(D + D_e) \frac{c^*}{\kappa y}.  
\end{equation}
The total pollutant flow rate is obtained by multiplying the flux by \(s\) and \(n\) to get
\begin{equation} \label{eq:mo_3}
    q_y = n s j_y \approx n s\left( v c - (D + D_e) \frac{c^*}{\kappa y}\right).   
\end{equation}
To express \(q_y\) in terms of \(c\), we use the fact that \eqref{eq:mo_0} can be evaluated at the average disturbance height $\bar{h}$ (m), to give
\begin{equation}
    c^* \approx -\frac{\kappa c_0}{\ln\left(\displaystyle\frac{\bar{h}}{y_0}\right)}.   
\end{equation}
Thus, \eqref{eq:mo_3} at $y = \bar{h}$ becomes
\begin{equation} \label{eq:mo_log_law}
    q_y \approx \frac{n s (D + D_e) c_0}{\bar{h} \ln(\bar{h} / y_0)}.  
\end{equation}
Both M-O theory and the log law assume fully developed turbulence, which may not fully capture irregular or transient turbulence caused by varying thermal conditions, weather patterns or vehicle disturbances.  

\subsubsection{Duct flow} \label{subsubsec:duct_flow}

For a fully-developed 2D channel flow centred around \(y=0\) with height \(2 H\), 
the velocity profile is described by \(u(y)\).
The pollutant flux to the channel wall is approximated by
\begin{equation}
    j_y = - D \frac{\partial c}{\partial y}(0) \sim \frac{D c}{h}. 
\end{equation}
Given that there are two walls and \(n\) channels, where \(s\) represents the surface area of each channel, the total pollutant flow rate to the channel walls can be expressed as
\begin{equation}
    q_y \sim \frac{n s (D + D_e) c}{h}.   
\end{equation}
Note that in systems with bends or varying cross-sections, additional complexities arise, which can affect the accuracy of the predictions.

\subsubsection{HVAC filters}

In this section, we derive the pollutant flow rates to the pleat walls and the fibre sheet in an HVAC filter, assuming the flow is fully developed.  
In Fig. 2$G$, the fibre sheet location varies linearly from the top to the bottom of the pleat.  
The velocity profile varies across the pleat channel and the velocity at the fibre sheet is given by $u_{\boldsymbol{n}}$.
The velocities in the duct ($u$) and the filter sheet ($u_{\boldsymbol{n}}$) can be connected using conservation of mass 
\begin{equation} \label{eq:mass_cons}
    \int_{A} u \, \text{d} A = \sum_{i=1}^m \int_{A_p} u_{\boldsymbol{n}} \, \text{d} A_p,
\end{equation}
where $A$ is the cross-sectional area of the duct, $A_p$ is the pleat area and there are \(m\) pleats in the HVAC filter.  
We match the average velocities to get
\begin{equation}
    \bar{u} A \approx m \bar{u}_{\boldsymbol{n}} A_p.
\end{equation} 
The pollutant flow rate to the pleat walls is driven by diffusion, as discussed in Section~\ref{subsubsec:duct_flow}.  
The rate of pollutant diffusion to the walls (area $s_w$, m$^2$) is expressed as
\begin{equation}
q_{y} \sim \frac{n m s_w (D + D_e) c}{h}.  
\end{equation}
The pollutant flow rate through the fibre sheet, \(q_{\boldsymbol{n}}\), is influenced by the collection efficiency \(E\).  
The collection efficiency is commonly assumed to have an exponential dependence on quantities such as the local velocity near the fibre sheet, the collection parameter and the diffusivity of the fibre material~\cite{hinds2022aerosol}.   
For simplicity, we impose the values found in Tables~\ref{tab:1a}--\ref{tab:1c}.  
The pollutant flow rate through the fibre sheet (area $A_p$, m$^2$) can be written as
\begin{equation}
q_{\boldsymbol{n}} \sim n m E A_p \bar{u}_{\boldsymbol{n}} c.  
\end{equation}
These formulas for pollutant flow rates to surfaces assume idealised conditions and do not account for real-world factors, such as airflow irregularities or particle variability, which may influence the actual performance of the systems.  

\subsection{Power, industry standards and drag measurements} \label{subsec:epandd}

\subsubsection{Natural flows}

\begin{figure}[t!]
    \centering
    \textsf{A} \hfill \textsf{B} \hfill \hfill \hfill \\
    \includegraphics[trim={2cm 1cm 2cm 3cm},clip,width=.49\linewidth]{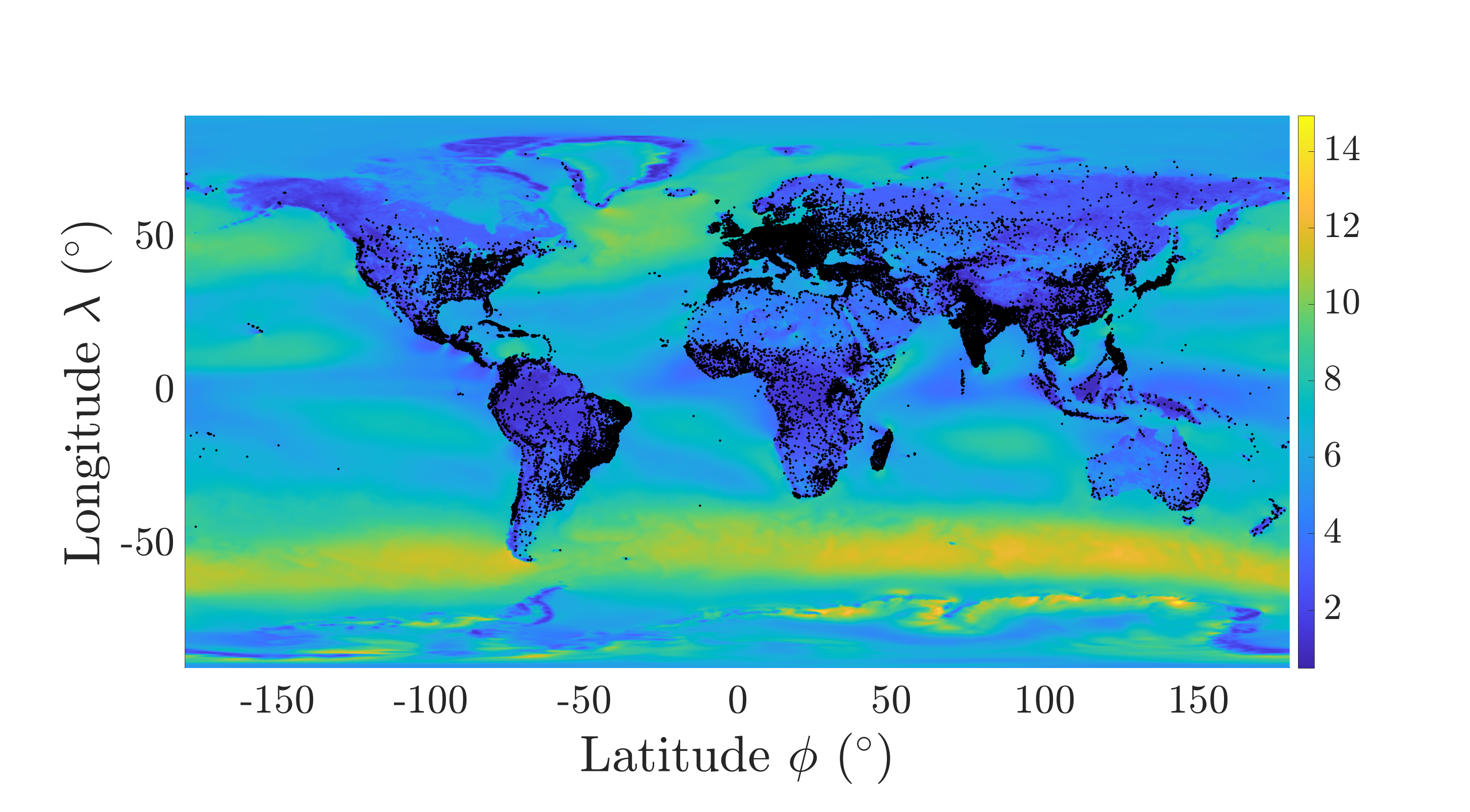} \hfill \includegraphics[trim={2cm 1cm 2cm 3cm},clip,width=.49\linewidth]{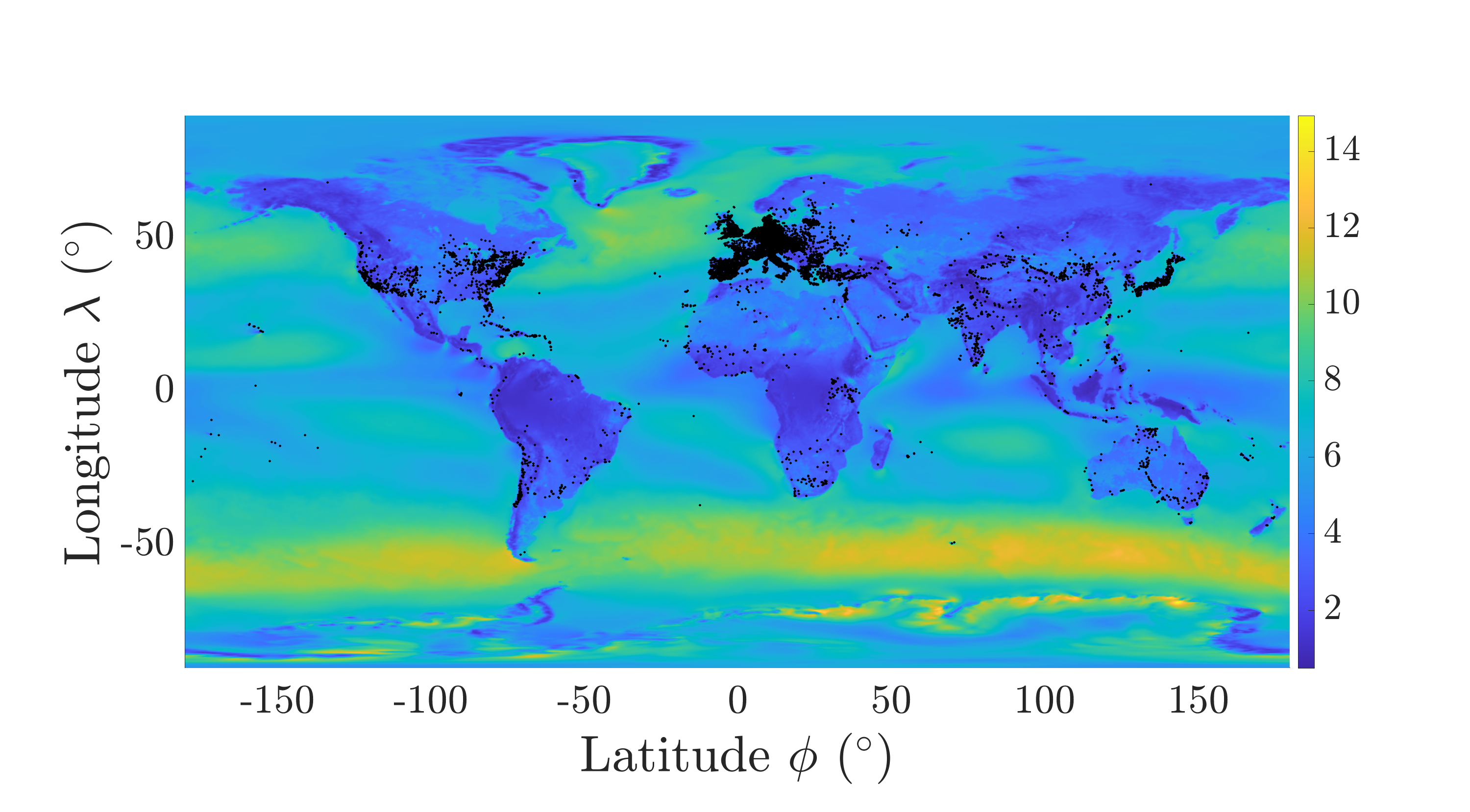} \hfill \hfill \hfill \\
    \textsf{C} \hfill \textsf{D} \hfill \hfill \hfill \\
    \includegraphics[trim={2cm 1cm 2cm 3cm},clip,width=.49\linewidth]{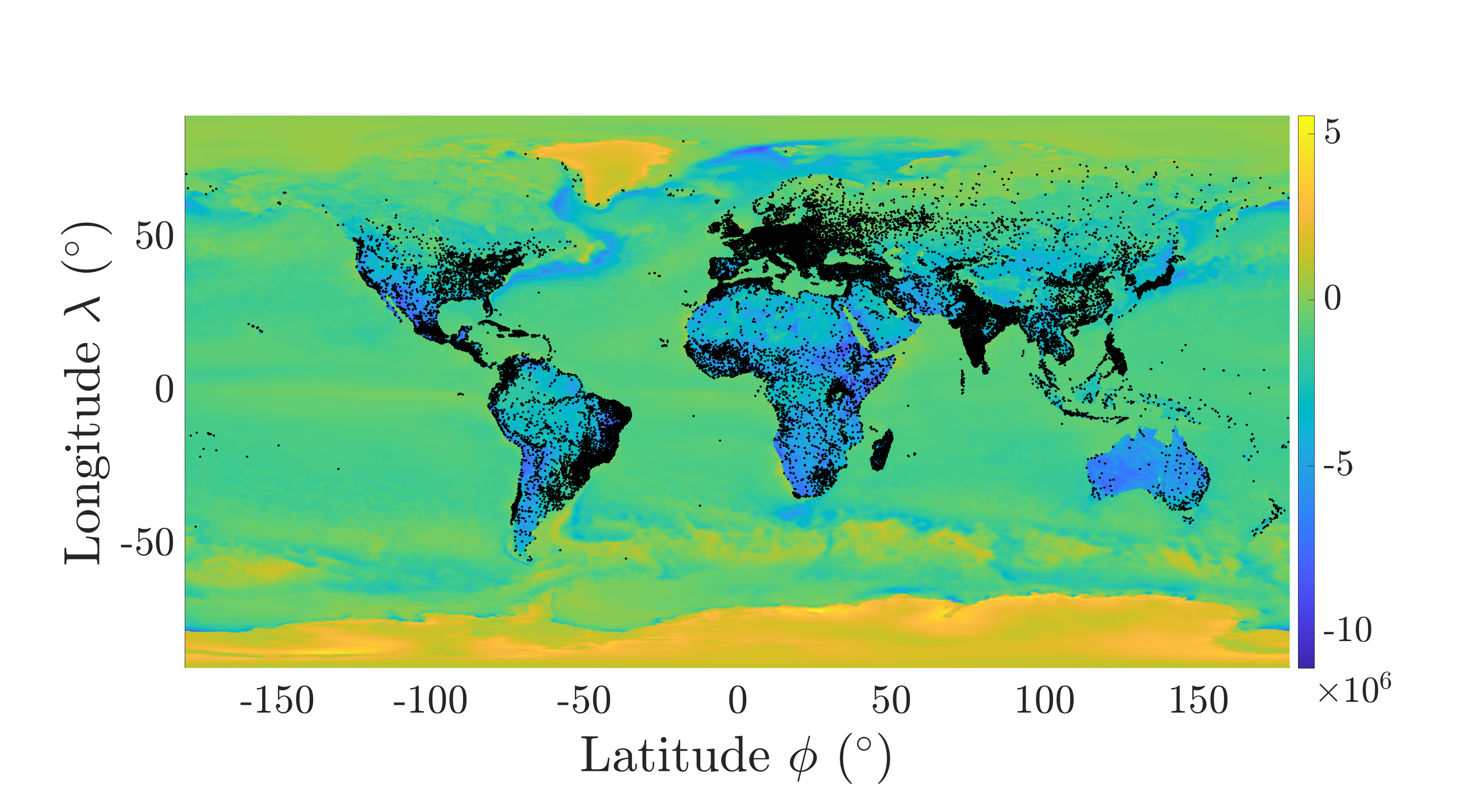} \hfill \includegraphics[trim={2cm 1cm 2cm 3cm},clip,width=.49\linewidth]{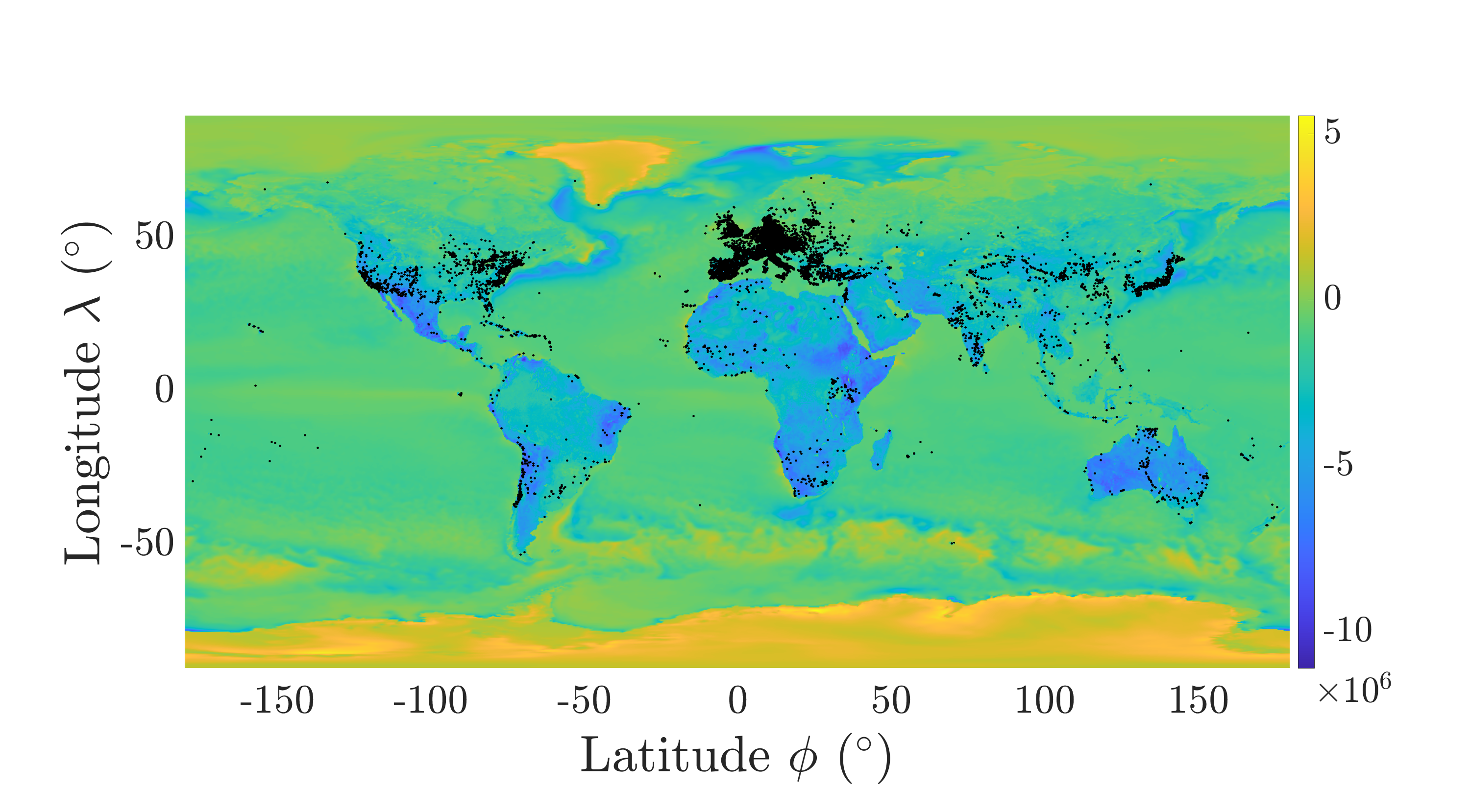} \hfill \hfill \hfill \\
    \caption{
    Contours of the global wind speed (m/s) distribution with the (A) world's cities in black and (B) world's solar farms in black, along with contours of the global sensible heat flux (W/m$^2$) distribution with the (C) world's cities in black and (D) world's solar farms in black~\cite{era5_single_levels, simplemaps_world_cities, dunnett2020harmonised}.  
    }
    \label{fig:natural_data}
\end{figure}

This section describes the relationships between \(Q\) and \( q_y \) in the context of heat transfer, specifically focusing on the effects of heat flux and temperature difference, as dictated by energy conservation.  
We consider a flat surface with an imposed heat flux \( \phi \) (W/m$^2$) over a specified area \( s = l^2 \) (m$^2$).  
The total energy entering the control volume due to this surface heat flux can be expressed as
\begin{equation} \label{eq:temp_1}
    \text{Energy in} = \phi s.  
\end{equation}
This represents the energy input to the system per unit time, driving thermal processes in the surrounding environment.  
Next, we analyse the energy exiting the control volume. 
This energy outflow can be characterised by the mass flow rate \( \dot{m} \) (kg/s), which is determined by the density \( \rho \) (kg/m$^3$), the free-stream velocity \( u \) (m/s) and the height of the control volume \( \delta \) (m)
\begin{equation} \label{eq:temp_2}
    \dot{m} = \rho u \delta l.  
\end{equation}
The energy leaving the control volume is then related to this mass flow rate and the specific heat capacity \( c_p \) (J/kgK) of the air, multiplied by the temperature difference \( \Delta T \) (K) between the surface and the surrounding atmosphere.   
This relationship can be expressed as
\begin{equation} \label{eq:temp_3}
    \text{Energy out} = \dot{m} c_p \Delta T = (\rho u \delta l) c_p \Delta T.  
\end{equation}
According to the principle of conservation of energy, we can equate the energy entering the control volume, \eqref{eq:temp_1}, to the energy exiting it, \eqref{eq:temp_2}--\eqref{eq:temp_3}, such that
\begin{equation} \label{eq:temp_4}
    \phi s = \rho u \delta l c_p \Delta T.  
\end{equation}
Rearranging \eqref{eq:temp_4} to isolate \( Q \sim u \delta l \), yields
\begin{equation} \label{eq:temp_5}
    Q = \frac{\phi s}{\rho c_p \Delta T}.  
\end{equation}
This expression provides a direct relationship between the streamwise flow rate and the physical parameters governing the heat transfer process.  
To derive an expression for the normal pollutant flow rate to the surface, \( q_y \), we utilise the scaling theory described in \eqref{eq:scaling}.
In order to relate \( q_y \) back to \(Q\), we multiply and divide by \(Q\) to get
\begin{equation} \label{eq:flux_relation}
    q_y \sim \frac{Q s (D + D_e) c}{\delta^2 u}.  
\end{equation}
Fig.~\ref{fig:natural_data} presents the variation in wind velocity and sensible heat flux across the world.  
Data for the wind velocity, boundary layer thickness, sensible heat flux and temperature is collected from ERA5~\cite{era5_single_levels}, with the world's cities and solar farms superimposed on top using SimpleMaps~\cite{simplemaps_world_cities} and Dunnett \textit{et al.}~\cite{dunnett2020harmonised} respectively.  
\eqref{eq:temp_5} provides a useful method for estimating airflow rates in scenarios influenced by thermal gradients, such as urban heat islands or solar energy collection.  
However, this approach may fail in highly turbulent conditions where the assumptions of steady-state flow do not hold~\cite{oke2002boundary}.

\subsubsection{Internal flows}

Duct and filter leakage can significantly affect the performance and energy efficiency of HVAC systems and HVAC filters~\cite{schild2009recommendations}.  
Typical duct leakage in HVAC systems ranges from 5\% to 20\%, depending on factors such as installation quality and system maintenance, with 10\% being a common assumption for well-maintained systems.
For HVAC filters, when properly installed, leakage is typically negligible and below 1\%.
However, poorly installed systems can experience leakage rates as high as 3\%.
To account for these losses in flow rate and energy calculations, a leakage factor should be applied.
This factor reduces the estimated airflow and adjusts energy consumption based on the system's specific leakage rate.
Incorporating these leakage factors ensures more accurate estimates of system performance and energy use. 
Note that these methods may be less accurate in situations with significant leakage or fluctuations in occupancy, complex building geometries or varying external conditions. 

\subsubsection{External flows}

For external flows over transportation systems, the streamwise flow rate \(Q\) can be estimated through the application of drag force dynamics~\cite{white2003fluid}. 
The drag force \( F_d \) (N) experienced by an object moving through a fluid is influenced by both the properties of the fluid and the object itself,
\begin{equation} \label{eq:ext_drag_1}
    F_d = \frac{1}{2} C_d \rho u^2 A,
\end{equation}
where \( C_d \) is the drag coefficient (--), \( \rho \) is the fluid density (kg/m\(^3\)), \( u \) is the average velocity of the object relative to the fluid (m/s) and \( A \) is the reference area (m\(^2\)) of the object exposed to the flow. 
To derive $Q$ in terms of $F_d$, we rearrange \eqref{eq:ext_drag_1} to incorporate \( Q = A u \), yielding
\begin{equation}
    F_d = \frac{C_d Q \rho u}{2}. 
\end{equation}
From this, the streamwise flow rate can be expressed as a function of the drag force acting on the transportation system
\begin{equation}
    Q = \frac{2 F_d}{\rho u C_d}. 
\end{equation}
The normal pollutant flow rate to the surface of the external flow can be evaluated using \eqref{eq:flux_relation}. 
Note that the above flow-rate formulas may fail to provide accurate estimates in scenarios involving complex flow patterns, such as turbulence, vortex shedding or crosswinds.

\renewcommand{\arraystretch}{1.4}
\begin{table*}[t!]
    \centering
    \caption{
    The flow type (--), the surface material (--), the density of the surface material (kg/m$^3$), the thickness of the surface material (m) and the flux of CO$_2$, CH$_4$ and N$_2$O (mol/kg/s) for different natural, internal and external flows~\cite{engineeringToolboxDensity, IBC2021, ASHRAE2024, SAE2024}.
    } 
    \vspace{0.5cm}
    \resizebox{\textwidth}{!}{
    \begin{tabular}{@{}lccccccc@{}}
    \toprule
    \text{Flow type} & \text{Material} & \text{Density (kg/m$^3$) } & \text{Thickness (m) } & Flux$_{\text{CO}_2}$ (mol/kg/s) & Flux$_{\text{CH}_4}$ (mol/kg/s) & Flux$_{\text{N}_2\text{O}}$ (mol/kg/s) \\
    \midrule
    \multirow{3}{*}{Natural flows} 
    & Concrete & 2400 & 0.1 & \(4 \times 10^{-10}\) & \(2 \times 10^{-12}\) & \(3 \times 10^{-13}\) \\
    & Glass & 2500 & 0.01 & \(1 \times 10^{-9}\) & \(6 \times 10^{-11}\) & \(1 \times 10^{-12}\) \\
    & Steel & 7850 & 0.005 & \(3 \times 10^{-10}\) & \(2 \times 10^{-12}\) & \(3 \times 10^{-13}\) \\
    \midrule
    \multirow{3}{*}{Internal flows} 
    & Aluminium & 2700 & 0.002 & \(1 \times 10^{-5}\) & \(6 \times 10^{-7}\) & \(8 \times 10^{-8}\) \\
    & Plastic & 950 & 0.005 & \(7 \times 10^{-6}\) & \(4 \times 10^{-7}\) & \(5 \times 10^{-8}\) \\
    & Stainless steel & 8000 & 0.003 & \(9 \times 10^{-7}\) & \(5 \times 10^{-8}\) & \(7 \times 10^{-9}\) \\
    \midrule
    \multirow{3}{*}{External flows} 
    & Aluminium & 2700 & 0.002 & \(2 \times 10^{-7}\) & \(1 \times 10^{-8}\) & \(2 \times 10^{-9}\) \\
    & Steel & 7850 & 0.005 & \(1 \times 10^{-8}\) & \(8 \times 10^{-10}\) & \(1 \times 10^{-10}\) \\
    & Composite & 1600 & 0.004 & \(3 \times 10^{-8}\) & \(2 \times 10^{-9}\) & \(2 \times 10^{-10}\) \\
    \bottomrule
    \end{tabular}}
    \label{tab:flux_per_material}
\end{table*}

\section{Average fluxes for surface materials}

Using Fig. 2, the average surface flux of each GHG was calculated for natural (\(\boldsymbol{j}_{\text{CO}_2}=1\times 10^{-7}\) mol/(m$^2$s), \(\boldsymbol{j}_{\text{CH}_4}=8\times 10^{-10}\) mol/(m$^2$s) and \(\boldsymbol{j}_{\text{N}_2\text{O}}=1\times 10^{-10}\) mol/(m$^2$s)), internal (\(\boldsymbol{j}_{\text{CO}_2}=1\times 10^{-4}\) mol/(m$^2$s), \(\boldsymbol{j}_{\text{CH}_4}=6\times 10^{-7}\) mol/(m$^2$s) and \(\boldsymbol{j}_{\text{N}_2\text{O}}=1\times 10^{-7}\) mol/(m$^2$s)) and external flows (\(\boldsymbol{j}_{\text{CO}_2}=2\times 10^{-6}\) mol/(m$^2$s), \(\boldsymbol{j}_{\text{CH}_4}=1\times 10^{-8}\) mol/(m$^2$s) and \(\boldsymbol{j}_{\text{N}_2\text{O}}=1\times 10^{-9}\) mol/(m$^2$s)). 
We convert from the surface flux per unit area (\(\boldsymbol{j}_{\text{pm}^2}\)) to the surface flux per unit mass (\(\boldsymbol{j}_{\text{pkg}}\)) for each surface material and flow type using the following relation
\begin{equation}
    \boldsymbol{j}_{\text{pkg}} = \frac{\boldsymbol{j}_{\text{pm}^2}}{\rho  t},
\end{equation}
where \(\rho\) is the density of the material (\(\mathrm{kg/m^3}\)) and \(t\) is the thickness of the material (\(\mathrm{m}\)) for either natural, internal or external flows.
Table~\ref{tab:flux_per_material} summarises the surface fluxes per unit mass for CO$_2$, CH$_4$ and N$_2$O. 
The surface fluxes of these GHGs per unit mass decrease from internal flows to external flows to natural flows, as well as when the mass per unit area \(\rho t\) increases across materials.

Concrete and aluminium are the most common materials used in these natural and mechanical flows. 
In the U.S., 93 million tons of concrete and 68.5 million tons of aluminium were produced in 2022, with the U.S. contributing about 10\% of global production. 
However, only a portion of this material is used in the exterior of buildings or HVAC systems. 
First, around 55\% of concrete is used in buildings, and of that, 25\% is used for exteriors. 
Second, approximately 25\% of aluminium is used in construction, with 7.5\% of that used in HVAC systems. 
Lastly, 27\% of aluminium is used in transportation, and of that, 45\% is used in transport exteriors, with 29\% of the aluminium used in transport exteriors being used in automobile exteriors. 
By evaluating these percentages of total produced materials and multiplying the surface flux of each GHG by the total mass produced per year, we get the produced flow rate of these GHGs to surfaces.
We approximate that 1 Gt/y of CO$_2$ is transported to natural surfaces, 0.1 Gt/y of CO$_2$ is transported to internal surfaces and 0.2 Gt/y of CO$_2$ is transported to external surfaces. 
We can compare these predictions to the other methods used to calculate the flow rate of these GHGs to produced surfaces in Fig.~\ref{fig:7}.

\section{The feasibility of different applications for surface-based atmospheric pollutant removal}

\subsection{Solar panels}

\updateii{Coating the surfaces of solar panels with sorbent or photocatalytic materials may reduce electrical output by partially occluding the panel surface. 
However, transparent TiO$_2$ coatings on solar panels have been demonstrated for NO$_\text{x}$ removal while largely preserving optical transmission~\cite{molar2025photocatalytic}, although their pollutant removal performance may differ from that reported for street and wall applications~\cite{Russell2021Review}. 
Solar farms are included primarily to illustrate potential atmospheric pollutant flow to large engineered surfaces, rather than as a practical strategy for atmospheric pollutant removal.} 

\subsection{HVAC systems}

\updateii{
HVAC systems are used here as representative high-airflow environments to estimate transport-limited pollutant delivery rates to surfaces. 
These calculations represent theoretical upper bounds on pollutant interception rather than a specific engineering design for catalytic integration within ventilation ducts. 
In ventilation environments, low-temperature technologies such as sorbents or photocatalysts would be more compatible with existing HVAC operation than high-temperature thermocatalysts. 
Photocatalytic systems require photon flux to drive oxidation reactions. 
UV irradiation is already deployed in some HVAC systems for microbial control, with characteristic doses of approximately 0.1 to 496 J/m$^{2}$ over typical exposure times of 0.1 to 16 s~\cite{thornton2022impact}, corresponding to irradiance levels of approximately 0.01 to 112 W/m$^{2}$. 
This range overlaps with the irradiance levels of 10 to 100 W/m$^{2}$ reported in photocatalyst studies for CH$_4$ oxidation~\cite{tsopelakou2024exploring}. 
However, practical implementation would depend on wavelength compatibility, residence time, catalyst efficiency, and the energy cost of UV illumination, none of which are captured in the transport-limited estimates presented here. 
}

\subsection{HVAC filters}

\updateii{HVAC filters progressively accumulate PM during operation, forming a filter cake that increases pressure drop and necessitates periodic replacement. 
Typical HVAC filters are replaced every 1 to 3 months depending on building type, outdoor air quality and operating conditions, with increasing PM loading raising fan energy demand prior to replacement. 
This periodic replacement is already standard practice in HVAC operation and could in principle provide an existing pathway for deploying sorbents or catalysts without introducing additional maintenance cycles; it may also help maintain catalytic or sorbent activity over time relative to static outdoor deployments.
If sorbents or catalysts are incorporated into HVAC filters~\cite{tomlinson2026modelling}, accumulated PM may block active sites and reduce removal efficiency. 
Sorbent saturation or catalyst deactivation may also occur independently of fouling, so material lifetime is governed by chemical and physical degradation mechanisms. 
For example, under quarterly replacement, \$600 per tCO$_2$ for a single installation cycle would increase to \$2400 per tCO$_2$ removed, with further increases for more frequent replacement or declining performance between cycles. 
Quantifying the coupled effects of filtration, fouling and material degradation is beyond the scope of this study but remains an important direction for future work. 
}

\subsection{Aeroplane wings and turbine blades}

\updateii{Surfaces such as wind turbine blades and aircraft wings are subject to strict aerodynamic and structural constraints; surface properties and mass distribution are critical to performance and lifetime, and any coating would need to satisfy these requirements. 
Wind turbine blades and aircraft wings are included primarily to illustrate atmospheric flow to surfaces, rather than to suggest practical pollutant removal at these surfaces.
} 

\section{O$_3$ using catalysis}

\update{Although our main analysis focused on CO$_2$, CH$_4$, PM$_{2.5}$ and NO$_\text{x}$, the methodology is general and can be applied to many atmospheric pollutants and removal technologies. 
One must specify the pollutant’s atmospheric concentration, diffusivity, surface area and airflow of the environment, calculating the atmospheric flow rate using Methods and applying the removal efficiency of the chosen technology. 
For example, O$_3$ is a short-lived climate pollutant linked to respiratory illness and crop yield losses~\cite{zhang2019ozone}, with typical atmospheric concentrations of 60 ppb ($3 \times 10^{-6}$ mol/m$^3$) and diffusivity of $2 \times 10^{-5}$ m$^2$/s. 
Using these values in \eqref{eq:empirical} for city environments (Tab. S2), gives an atmospheric flow rate to all city surfaces of around $3 \times 10^{3}$ mol/s. 
With a photocatalytic removal efficiency of 0.5 given by Ohtani et al.~\cite{ohtani1992} for TiO$_2$, the above flow rate corresponds to a removal potential of $5\times10^{-3}$ GtO$_3$/y, representing approximately
1\% of the tropospheric O$_3$ burden of around 0.4 Gt~\cite{zhang2019ozone}. 
}

\updateii{A relevant real-world example of catalytic pollutant removal is the PremAir™ radiator coating deployed in certain vehicles in the 1990s~\cite{greger1998premair}.
This system incorporated a catalyst into the vehicle radiator to decompose ambient O$_3$ as air passed through the cooling airflow. 
Laboratory experiments reported O$_3$ removal efficiencies of up to 90\%. 
However, fleet-scale analyses concluded that the resulting reduction in urban O$_3$ concentrations was modest because the catalytic surface area and airflow rate represented only a small fraction of the atmospheric volume, and deployment was difficult to justify without strong regulatory incentives~\cite{greger1998premair}.}

\begin{figure*}[t!] 
\centering 
\includegraphics[width=\linewidth]{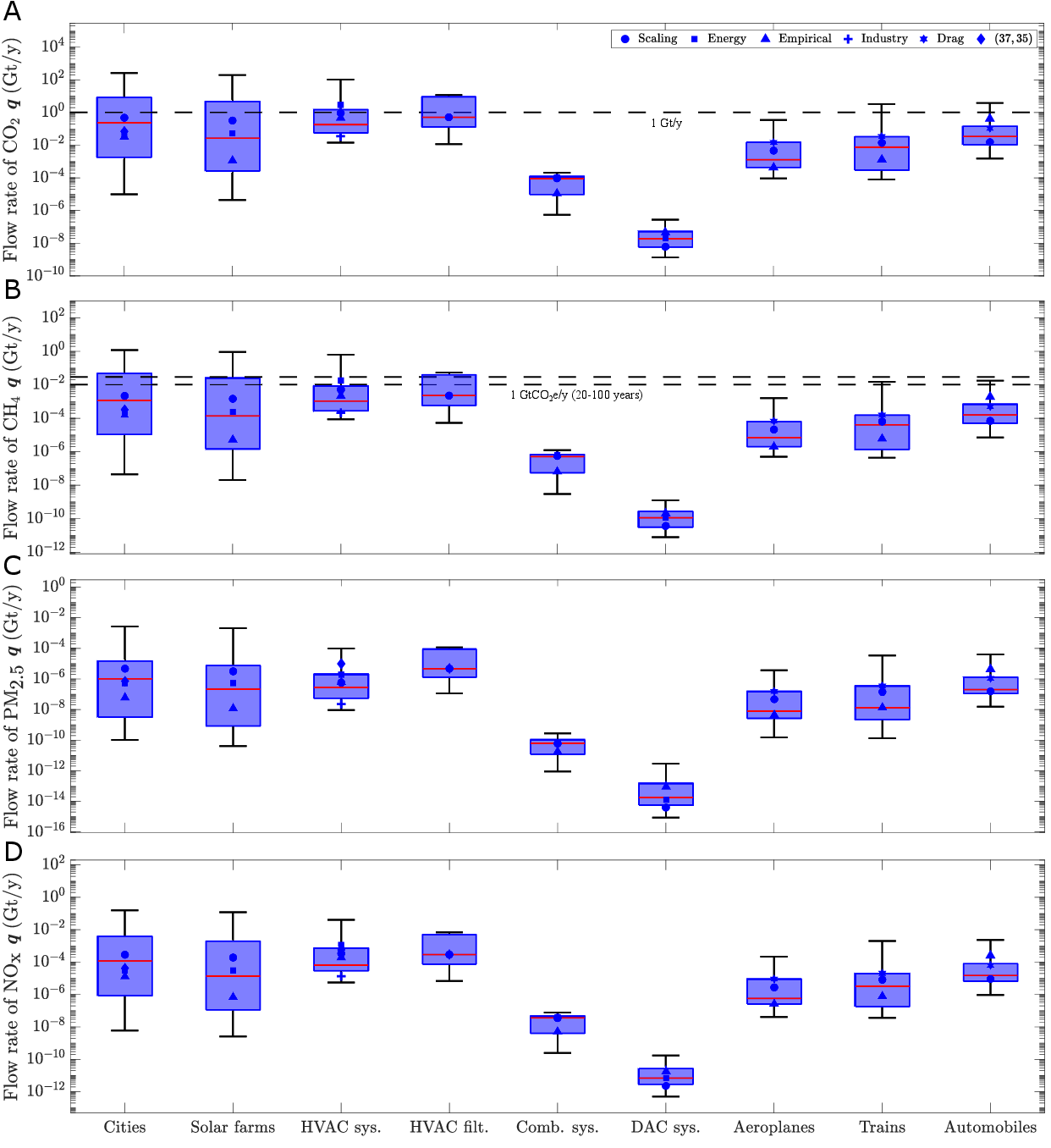} 
\caption{
Flow rates ($\boldsymbol{q}$) of atmospheric (A) CO$_2$, (B) CH$_4$, (C) PM$_{2.5}$ and (D) NO$_\text{x}$ to the annually produced surfaces in natural (e.g., cities, solar farms), internal (e.g., HVAC, combustion, DAC systems) and external (e.g., aeroplanes, trains, automobiles) environments, with symbols representing the methods used to estimate flow rates.
} 
\label{fig:7} 
\end{figure*} 

\section{Produced flow rate}

In Fig.~\ref{fig:7}, we present the flow rate of atmospheric CO$_2$, CH$_4$, PM$_{2.5}$ and NO$_\text{x}$ to the produced surfaces of cities, solar farms, HVAC systems, combustion systems, DAC systems and aeroplanes, trains and automobiles.
This figure can be compared with the flow rate of atmospheric pollutants to the total surfaces in Fig. 3 of the main text.

\section{Other climate metrics}

\updateii{Finally, we note that the 100-year GWP remains a commonly used regulatory and corporate reporting metric. 
Under the 100-year GWP, methane's CO$_2$-equivalent factor is approximately 27 to 30, compared with approximately 80 to 84 under the 20-year GWP~\cite{ipcc2023synthesis}. 
Consequently, the CO$_2$e benefit per tonne of CH$_4$ potentially removed would be approximately three times lower than the values reported here, and the cost per tCO$_2$e potentially removed would increase proportionally. 
Alternative metrics may better represent specific policy or temperature-based accounting objectives. 
While the choice of equivalency metric affects the interpretation of CO$_2$e benefits, it does not alter the atmospheric pollutant fluxes to surfaces estimated in this study. 
}

\bibliography{achemso-demo_2}